\documentclass[12pt]{article} 
\usepackage{epsfig}
\usepackage{amsfonts}
\usepackage{latexsym}
\usepackage{amsmath,amssymb}
\usepackage{verbatim}
\usepackage{setspace}
\usepackage[small]{caption}
\usepackage{graphicx}

\usepackage{color}
\definecolor{darkgreen}{rgb}{0,0.5,0}
\definecolor{darkblue}{rgb}{0,0,0.6}
\definecolor{purple}{rgb}{0.4,.2,0.7}
\usepackage[colorlinks=true,citecolor=darkgreen,linkcolor=purple,urlcolor=purple]{hyperref}
\usepackage[textheight=9in, textwidth=6.5in, letterpaper]{geometry}

\numberwithin{equation}{section}

\def\tilde{\widetilde}
\newcommand{\bea}{\begin{eqnarray}}
\newcommand{\eea}{\end{eqnarray}}
\newcommand{\be}{\begin{equation}}
\newcommand{\ee}{\end{equation}}
\newcommand{\ba}{\begin{align}}
\newcommand{\ea}{\end{align}}

\renewcommand{\epsilon}{\varepsilon}

  \makeatletter
  \let\over=\@@over \let\overwithdelims=\@@overwithdelims
  \let\atop=\@@atop \let\atopwithdelims=\@@atopwithdelims
  \let\above=\@@above \let\abovewithdelims=\@@abovewithdelims

\renewcommand\section{\@startsection {section}{1}{\z@}%
                                   {-3.5ex \@plus -1ex \@minus -.2ex}
                                   {2.3ex \@plus.2ex}%
                                   {\normalfont\large\bfseries}}

\renewcommand\subsection{\@startsection{subsection}{2}{\z@}%
                                     {-3.25ex\@plus -1ex \@minus -.2ex}%
                                     {1.5ex \@plus .2ex}%
                                     {\normalfont\bfseries}}

\def\Or[#1]{{\text{O}}\left({#1}\right)}
\def\dotl[#1,#2]{\left\langle #1, #2 \right\rangle}
\def\dotlb[#1,#2]{[ #1, #2 ]}
\def\dotp[#1,#2]{(#1) \cdot (#2)}
\def\aff[#1,#2]{\hat{#1}(#2)}
\def\n4sym{{\cal N}=4 SYM}
\def\>{\rangle}
\def\<{\langle}
\def\weight[#1,#2,#3]{\{(#1),#2,#3\}}
\def\ads[#1]{$\text{AdS}_{#1}$}

\begin{document}
\unitlength = 1mm
\ \\

\begin{center}

{ {\LARGE {\textsc{The Wavefunction of Vasiliev's Universe }} } \\ {  {\textsc{ (A Few Slices Thereof)}}}}

\vspace{0.8cm}

Dionysios Anninos$^1$, Frederik Denef$^{2,3}$ and Daniel Harlow$^1$

\vspace{1cm}

\vspace{0.5cm}

$^1$ {\it Stanford Institute for Theoretical Physics, Stanford University}\\

$^2$ {\it Institute for Theoretical Physics, University of Leuven}\\

$^3$ {\it Center for the fundamental Laws of Nature, Harvard University}

\vspace{1.0cm}

\end{center}

\begin{abstract}

We study the partition function of the free $Sp(N)$ conformal field theory recently conjectured to be dual to asymptotically de Sitter higher-spin gravity in four-dimensions. We compute the partition function of this CFT on a round sphere as a function of a finite mass deformation, on a squashed sphere as a function of the squashing parameter, and on an $S^2 \times S^1$ geometry as a function of the relative size of $S^2$ and $S^1$.  We find that the partition function is divergent at large negative mass in the first case, and for small $S^1$ in the third case.  It is globally peaked at zero squashing in the second case.  Through the duality this partition function contains information about the wave function of the universe.  We show that the divergence at small $S^1$ occurs also in Einstein gravity if certain complex solutions are included, but the divergence in the mass parameter is new.  We suggest an interpretation for this divergence as indicating an instability of de Sitter space in higher spin gravity, consistent with general arguments that de Sitter space cannot be stable in quantum gravity.  



\end{abstract}

\pagebreak 
\setcounter{page}{1}
\pagestyle{plain}

\setcounter{tocdepth}{1}

\tableofcontents
\section{Introduction}

The dS/CFT correspondence is perhaps the most well-known program for incorporating cosmologically relevant spacetimes into some form of holographic theory \cite{Strominger:2001pn,Hull:1998vg,Witten:2001kn,Strominger:2001gp,Maldacena:2002vr}.\footnote{For other proposals see \cite{Alishahiha:2004md,Dong:2010pm,Dong:2011uf,Banks:2011qf,Banks:2008ep,Freivogel:2006xu,Harlow:2012dd,Garriga:2008ks,Parikh:2004wh,Anninos:2011af}.}  One proposal for the dS/CFT dictionary\footnote{For a recent review on some aspects of asymptotically de Sitter spaces see \cite{Anninos:2012qw}.} \cite{Maldacena:2002vr} is that the Hartle-Hawking wave function \cite{PRINT-83-0937 (CAMBRIDGE)} in the bulk is equal to the partition function of a Euclidean CFT deformed by various operators with finite coefficients (see also \cite{Maldacena:2011mk,Harlow:2011ke,Hertog:2011ky,Anninos:2011kh,{Castro:2011xb},McFadden:2009fg,Anninos:2010zf,arXiv:1106.1175}).  In this paper we study that proposal in a recent explicit realization of the dS$_4$/CFT$_3$ correspondence \cite{arXiv:1108.5735}, in which the bulk is Vasiliev's higher spin gravity \cite{Vasiliev:1990en,Vasiliev:1999ba} in dS$_4$ \cite{Iazeolla:2007wt,Vasiliev:1986td}.  We explicitly compute the partition function with some finite deformations turned on and discuss the implications for the wave function of Vasiliev's universe.

The duality of \cite{arXiv:1108.5735} was inspired by similar efforts in the context of AdS$_4$ higher spin gravity \cite{Klebanov:2002ja,Witten,Mikhailov:2002bp,Sezgin:2002rt,hep-th/0103247,arXiv:0912.3462,arXiv:1102.2219}. The dual CFT is a (Euclidean) free theory of anti-commuting scalar fields $\chi^a$, with flat space Lagrangian \cite{LeClair:2006kb,LeClair:2006mb,LeClair:2007iy}
\begin{equation}
\label{spinlag}
\mathcal{L}_{CFT} =  \frac{1}{2}\Omega_{ab} \partial_i \chi^a \partial^i \chi^b~.
\end{equation}
Here $\Omega_{ab}$ is the symplectic form, with $a,b \in \{ 1,2,\ldots,N\} $, and the $\chi^a$ transform as an $Sp(N)$ vector.\footnote{Note that in our conventions $N$ must be even.}  Repeated indices are summed over.  $N$ is related to the bulk cosmological constant $\Lambda_{bulk} \equiv 3/\ell_{dS}^2$ as $N \sim 1/(G\Lambda_{bulk})$.

To get a bulk spectrum that matches Vasiliev's, the operator content is restricted to the $Sp(N)$ singlet sector.  This is built from a set of even-spin bilinear currents of the general form $\Omega_{ab}\chi^a \partial_{\mu_1}\ldots \partial _{\mu_s} \chi^b$, where the derivatives can act either to the left or the right and are symmetrized with various traces removed such that for $s\geq2$ the current is symmetric, traceless, and conserved.  We will be especially interested in the spin zero `current'
\begin{equation}
J^{(0)} = -\frac{1}{2}\Omega_{ab} \chi^a \chi^b~,
\end{equation}
which has conformal dimension one, and the spin two current $T_{ij}$, which is the conserved traceless stress tensor \cite{Callan:1970ze}\footnote{These operators are renormalized in a way we will make explicit below.}
\be\label{stresstensor}
T_{ij}=\frac{1}{4}\Omega_{ab}\left[-3\partial_i\chi^a\partial_j \chi^b+\chi^a\partial_i\partial_j\chi^b+\delta_{ij}\partial_k \chi^a \partial^k \chi^b\right].
\ee
This stress tensor has conformal dimension three. 

$J^{(0)}$ is dual to a bulk scalar field $\phi$ with mass $m_{bulk}^2\ell_{dS}^2 = 2$.  $T_{ij}$ is dual to the bulk graviton.  More generally, the spin $s$ current is dual to a massless spin $s$ field in the bulk.  In relating the bulk and boundary there is a subtlety for the scalar $\phi$ in that the dimension of $J^0$ is less than $3/2$, so what is usually called the ``alternate quantization'' in AdS/CFT is in play here.  Turning on a field theory source $\sigma$ for $J^0$ does not correspond to Dirichlet boundary conditions for the bulk scalar field $\phi$ at the future boundary; as we explain in appendix \ref{altquantapp} it instead corresponds to fixing a linear combination of $\phi$ and its canonical momentum $\Pi$ at the late-time cutoff surface.

To be explicit, recall the dS/CFT dictionary in \cite{Maldacena:2002vr}:
\be
\Psi_{HH}[\epsilon^{d-\Delta}\tilde{\sigma},\epsilon^{-2}g_{ij}]=\int\mathcal{D}\Phi \,e^{-S_{CFT}[g_{ij},\Phi]+\int d^d x \sqrt{g}\tilde{\sigma}\mathcal{O}[\Phi]},
\ee
where $\Psi_{HH}[\phi,g]$ is a Wheeler-de Witt wave function for the scalar field configuration at the boundary of spacetime to be $\phi$ and the induced metric on the boundary to be $g$ computed from an integral over compact bulk geometries with no other boundaries.  The equality really is meant asymptotically as the ``cutoff'' parameter $\epsilon$ goes to zero.  $\Delta$ is the dimension of the dual operator $\mathcal{O}$, and $S_{CFT}$ is understood to include divergent local terms in the sources $\tilde{\sigma}$ and $g$, which are tuned to match the divergent local pieces of $\Psi_{HH}$.\footnote{It is somewhat unsatisfying that the scheme-dependent local terms in the field theory partition function must be determined by comparison to the bulk, but the fact is that the local terms in the wave function do contain physical information.  From the field theory point of view there is not any particular reason to choose one scheme over another, with the exception of trying to subtract the divergences altogether.  This choice of scheme is incorrect here, since the divergences really do exist on the bulk side.  One might hope that it is possible to formulate a purely field theoretic condition which determines the ``correct'' coefficients for divergent local terms in any particular cutoff scheme.}  In the alternate quantization instead of fixing the value of $\phi$ at the boundary we instead fix a quantity which for the $Sp(N)$ model is 
\be\label{altBC}
\phi-\epsilon^3 \Pi=\frac{1}{4}\sqrt{N}\epsilon^2\sigma~.
\ee
Here $\Pi=-i\frac{\delta}{\delta\phi}$ is the momentum conjugate to $\phi$ acting on the wave function.  In Fefferman-Graham gauge, where the metric approaches $ds^2={(\ell_{dS}/T)^2}\left({-dT^2+g_{ij} dx^i dx^j}\right)$ as $T\to0^-$, we have
\be
\Pi=T^{-2}\partial_T \phi~.
\ee
The coefficients of $\phi$ and $\Pi$ in \eqref{altBC} are chosen to ensure conformal invariance with the correct dimension for the dual operator $\mathcal{O}$.  We can thus interpret the free $Sp(N)$ partition function as the same Hartle-Hawking wave function projected onto a basis of eigenstates of the Hermitian operator $\phi-\epsilon^d \Pi$.  Unlike in AdS/CFT there is only one bulk quantization; changing the scalar boundary conditions is simply changing the quantum mechanical basis for the wave function.  For more details on this point see appendix \ref{altquantapp}.

In AdS/CFT it is usually sufficient to restrict attention to the partition function with arbitrary infinitesimal sources turned on for the boundary operators; for us this would have the form
\begin{equation}\label{linearZ}
Z_{CFT} [\sigma,g_{ij},\ldots] = \int \mathcal{D} {\chi} e^{- \int d^3 x \left[  \mathcal{L}_{CFT} - \sigma(x) J^{(0)} + \frac{1}{2}T^{ij} (x) h_{ij} + \ldots \right] }~,
\end{equation}
where $\ldots$ are linear couplings to the higher spin currents and also any counterterms.  

This linearized expression allows computation of bulk correlators with fixed future boundary conditions $\sigma=0$, $h_{ij}=0$ \cite{Harlow:2011ke,Anninos:2010zf,arXiv:1106.1175}, but in order to make sense of the full wave function proposal
\begin{equation}
\label{wfdict}
\Psi_{HH} \left[ \epsilon^2 \sigma,\epsilon^{-2}g_{ij},\ldots \right] = Z_{CFT} [\sigma,g_{ij},\ldots]~,
\end{equation}
we need to understand how to generalize \eqref{linearZ} to finite values of $\sigma$, $h_{ij}$, etc.  In this paper we will consider only nonzero $\sigma$ and $h_{ij}$, since these couple to relevant and marginal operators in the Wilsonian sense.  Intuitively $\sigma$ gives a position-dependent mass term and $h_{ij}$ introduces a background metric in the CFT.  The higher spin currents are irrelevant, and turning them on at finite strength introduces considerable UV subtlety into the definition of the partition function.  The presence of higher-spin symmetry may constrain these operators in a strong enough way that the partition function remains well-defined despite their irrelevance, but we will not attempt to demonstrate this here.  For the most part we will just set all of the higher-spin sources to zero.\footnote{A point that has been emphasized by Steve Shenker is that single-trace irrelevant operators do not appear in conjectural duals to more conventional dS theories that do not have higher spin massless fields in the bulk.  In particular the operator dual to a scalar with positive mass squared never has a real part which is greater than $d$, and the operators dual to gauge fields and the graviton are always at most marginal.}

Our interpretation of this wave function will be the same as that of Hartle and Hawking \cite{PRINT-83-0937 (CAMBRIDGE)}, which is that its square computes conditional probabilities.  So for example
\be
P[\sigma,g_{ij}]\equiv|\Psi_{HH}[\epsilon^2 \sigma,\epsilon^{-2}g_{ij},0,0,\ldots]|^2
\ee
gives the probability that the induced metric on some spatial surface is $g_{ij}$ and that the scalar and its canonical momentum obey \eqref{altBC}, given that the higher spin fields are all zero.
 
The outline for the remainder of the paper is as follows: in section \ref{s3section} we introduce a renormalization procedure for dealing with finite sources and use it to compute the partition function on a round $S^3$ as a function of a constant source $\sigma$ for $J^{(0)}$.  We discover that it diverges at large negative $\sigma$.  In section \ref{sqs3sec} we compute the partition function of the $Sp(N)$ model on a squashed three sphere as a function of the squashing.  We find that it is globally peaked at zero squashing.  In section \ref{s2s1sect} we compute the partition function on $S^2\times S^1$ as a function of the relative size of $S^2$ and $S^1$, finding that it diverges at small $S^1$.  In section \ref{einstsec} we recall the bulk wave function of a free massive scalar and observe that there is no divergence of the type found in section \ref{s3section}.  We then do an Einstein gravity dS version of the Hawking-Page calculation of the wave function on $S^2\times S^1$, finding that we are able to produce a divergence of the same type as we found from the dual field theory in section \ref{s2s1sect}.  In section \ref{secmore} we return to the scalar divergence; we show that it implies that the critical version of the $Sp(N)$ model at finite $N$ does not exist on $S^3$. We interpret this as an instability of higher spin gravity, modulo a certain technical point about integrating over the higher-spin fields that we are unable to decisively resolve.  In section \ref{bigbang} we briefly point out a relationship between the $Sp(N)$ model on quotients of $\mathcal{H}_3$ and bulk `big bang' cosmologies. Section \ref{outlook} is our conclusion, and in appendix \ref{altquantapp} we give a detailed treatment of double-trace flow in dS/CFT and the dS interpretation of the ``alternate quantization''.




\section{Finite Deformations and the Wave Function on the Three Sphere}\label{s3section}

The obvious guess for how to turn on finite $h_{ij}$ and $\sigma$ is to just put the theory \eqref{spinlag} on a background $g_{ij}$ and turn on a position dependent mass $\sigma$.  Including the conformal mass term this gives
\be
\label{sourceact}
S_{bare}[\sigma,g_{ij},\chi]=\frac{1}{2}\int d^3x\sqrt{g}\Omega_{ab}\left[\partial_i\chi^a \partial _j \chi^b g^{ij}+\frac{1}{8}R[g]\chi^a\chi^b+\sigma(x)\chi^a\chi^b\right].
\ee
Justifying this expression however involves some care.  In particular putting the theory on a background metric $g_{ij}$ involves nonlinear terms in the perturbation $h_{ij}$; for example the conformal mass is required to preserve tracelessness of the stress tensor.  One might worry that nonlinear terms in $\sigma$ similarly need to be included in the bare action of the field theory in order to properly reproduce the bulk physics, or even that terms involving higher powers of the curvature tensor and its derivatives need to be included.\footnote{We thank Dan Jafferis and Douglas Stanford for discussions of this issue.}  Fortunately in the free $Sp(N)$ model any term involving $\sigma^2$ is always irrelevant in the Wilsonian sense, as are most things involving the curvature tensor.  The only relevant (or marginal) terms that can be added depend only on the sources:
\be\label{counterterms}
S_{loc}=A\Lambda^3\int d^3 x \sqrt{g}+B\Lambda\int d^3 x \sqrt{g}R[g]+C\Lambda\int d^3x\sqrt{g}\sigma,
\ee
where $\Lambda$ is the energy cutoff.  We are here assuming that the theory is cutoff in a geometric way so that all terms are diffeomorphism invariant.  The linear term in $\sigma$ has a simple interpretation as a freedom to shift the identity part of the composite operator $\chi^2$, and the other terms are renormalizations of the energy momentum tensor.  There is also a relevant operator $(J^{(0)})^2$ and a marginal operator $(J^{(0)})^3$, but these must have zero coefficient to stay in the free theory in the IR when $\sigma=0$.

The need to impose the singlet constraint also leads to an additional subtlety, which is typically dealt with by coupling the theory to Chern-Simons theory and taking the limit $k\to\infty$.  Fortunately this does not affect the partition function on simply-connected manifolds like $\mathbb{R}^3$ and $S^3$, since there are no nontrivial flat connections, so we will postpone further discussion of the Chern-Simons sector until section \ref{s2s1sect}.\footnote{A point that has been emphasized to us by Simeon Hellerman is that on $S^3$ the pure Chern-Simons partition function vanishes in the limit $k\to\infty$.  Since we should really multiply \eqref{Zdet} below by this partition function, the result becomes trivial.  For this reason we have to imagine studying the theory at large but finite $k$, as emphasized in \cite{shenkermaltz}.  This vanishing however cancels out of any relative probabilities computed on the topological three-sphere.  In this paper we will always compare different metrics and scalar field configurations on the same topology, either the three sphere or $S^2\times S^1$, so our results for these probabilities will be excellent approximations to the ``true'' finite but large $k$ theory.}  

In the cases where we may neglect the Chern-Simons sector, we may formally evaluate the path integral:
\be
\label{Zdet}
Z_{CFT}[\sigma,g_{ij}]=e^{-S_{loc}}\int \mathcal{D}\chi e^{-S_{bare}[\sigma,g_{ij},\chi]}=e^{-S_{loc}}\det \left(\frac{-\nabla^2+\sigma+\frac{1}{8}R[g]}{a^2\Lambda^2}\right)^{N/2}.
\ee
Here we have defined the path integral measure as
\be
\mathcal{D}\chi^a=\prod_n \frac{dc_n^a}{a\Lambda},
\ee
where $c_n^a$ are the Grassman coefficients of $n$-th eigenfunction of the conformal Laplacian $-\nabla^2+\frac{1}{8}R[g]$ in the mode expansion of $\chi^a$, and $a$ is a dimensionless coefficient which can be adjusted by rescaling $\chi$ and is thus a possible wavefunction renormalization.

The coefficients of the local terms in $S_{loc}$ are in principle determined by matching to a bulk calculation of the wave function, but in Vasiliev gravity this is beyond reach since the action is not known.  We will instead observe that for typical bulk theories the divergences in $\Psi_{HH}[\phi,g]$ are pure phase, so that they cancel in the probability $\Psi\Psi^*$.  For example we'll see this is true for pure Einstein gravity or a free bulk scalar in section \ref{einstsec} below.  The general reason is that if the bulk Lorentzian action is real, the dominant contribution to the action at large spatial boundary comes from real classical field configurations near the boundary which produce a phase $e^{iS}$.  Without the action we cannot be completely sure that this works in Vasiliev theory, but in any case it is required for there to be a well-defined de Sitter invariant probability distribution $|\Psi_{HH}|^2$ at late times. From here on we will assume that the divergent local pieces in $\Psi_{HH}[\phi,g]$ are indeed pure phase.\footnote{This can also be justified by analytic continuation from the AdS version of Vasiliev's theory if we assume that that all local terms are real in a Euclidean AdS calculation of the partition function as a function of boundary sources.  By dimensional analysis the coefficients of these terms are always an odd power of $\ell_{ads}$, divided by $G$, and multiplied by a power series in $G \ell_{ads}^{-2}$. Continuing $\ell_{ads}\to i \ell_{ds}$ then gives local terms which are purely imaginary for real sources.}   We can thus define new partition function $Z_{finite}$, which is related to $Z_{CFT}$ by a divergent local phase, by
\be\label{Zdetfin}
Z_{finite}[\sigma,g_{ij}]=e^{-\hat{S}_{ct}}\det \left(\frac{-\nabla^2+\sigma+\frac{1}{8}R[g]}{a^2\Lambda^2}\right)^{N/2}.
\ee
Here $\hat{S}_{ct}$ is of the same form as \eqref{counterterms}, but with $\hat{A}$, $\hat{B}$, $\hat{C}$, and $a$ now chosen to cancel the UV divergences of the determinant.  This new definition is extremely convenient since it can be unambiguously implemented on the field theory side, and with our phase assumption we have
\be
|\Psi_{HH}|^2=|Z_{CFT}|^2=|Z_{finite}|^2
\ee
so we may still extract the probability distribution from $Z_{finite}$.

\subsection{Partition Function with a Constant Mass}\label{sig0sec}
We first discuss the simple case of a constant mass deformation $\sigma_0$ on a round $S^3$.  We will take a simple hard cutoff in the sum over modes:
\be\label{hardcutoffsum}
2\log Z_{finite}=-2\hat{S}_{ct}+N\sum_{\ell=0}^{\ell_{max}} (\ell+1)^2\log \left(\frac{\ell(\ell+2)+\frac{3}{4}+\sigma_0}{a^2\Lambda^2}\right).
\ee
We have used that the spectrum of the Laplacian on $S^3$ is $-\ell(\ell+2)$ for $\ell=0,1,2,\ldots$, that the degeneracy is $(\ell+1)^2$, and have chosen units where the sphere radius is 1.  To ensure that the cutoff is geometric, we insist that it is really on eigenvalues of the conformal Laplacian:
\be \label{cutoff}
\ell_{max}(\ell_{max}+2)+\frac{3}{4}=\Lambda^2.
\ee
First consider the case where $\sigma_0=0$.  The sum can be done in Mathematica, and its asymptotic expansion for large $\Lambda$ is
\begin{align}\nonumber
2N^{-1}\log Z_{finite}=&-2N^{-1}\hat{S}_{ct}-\frac{2}{9}(1+3\log a)\Lambda^3-\log a \, \Lambda^2-\frac{1}{12}(1+7\log a)\Lambda\\
&+\frac{1}{8}\left(\log 4-2\log a-\frac{3 \zeta(3)}{\pi^2}\right)+O(1/\Lambda). 
\end{align}
Cancellation of the quadratic divergence requires $a=1$, and cancellation of the cubic and linear divergences fixes $\hat{A}=-\frac{N}{18\pi^2}$ and $\hat{B}=-\frac{N}{188\pi^2}$.  We thus find 
\be\label{finite0mass}
\log Z_{finite}=\frac{N}{16}\left(\log 4-\frac{3\zeta(3)}{\pi^2}\right),
\ee
which is the continuation $N\to-N$ of the well-known result in the $O(N)$ model \cite{arXiv:0908.2657,arXiv:1105.4598}.  If we had not defined $\Lambda$ in a geometric way as in \eqref{cutoff} we would have gotten the wrong finite part.

Turning the mass $\sigma_0$ back on and imposing the cutoff as
\be \label{cutoffmass}
\ell_{max}(\ell_{max}+2)+\frac{3}{4}+\sigma_0=\Lambda^2,
\ee
a similar argument sets $\hat{C}=\frac{N}{4\pi^2}$ and gives
\be
2\log Z_{finite}=\frac{N}{8}\left(\log 4-\frac{3\zeta(3)}{\pi^2}\right)+N\sum_{\ell=0}^{\infty}(\ell+1)^2\left[\log\left(1+\frac{\sigma_0}{\ell(\ell+2)+\frac{3}{4}}\right)-\frac{\sigma_0}{\ell(\ell+2)+\frac{3}{4}}\right].
\ee
The sum on the right-hand side can be done by differentiating with respect to $\sigma_0$, performing the sum, and then integrating.  The constant can be fixed by matching to \eqref{finite0mass} when $\sigma_0=0$.  The final result is
\begin{align}\nonumber
\log Z_{finite}=\frac{N}{48\pi^2}\Big[&6\pi^2(1-4\sigma_0)\log\left(1-e^{-i\pi\sqrt{1-4\sigma_0}}\right)+12\mathrm{Li}_{3}\left(e^{-i\pi\sqrt{1-4\sigma_0}}\right)\\
&+i\pi\sqrt{1-4\sigma_0}\left(\pi^2(1-4\sigma_0)+12 \mathrm{Li}_2 \left(e^{-i\pi\sqrt{1-4\sigma_0}}\right)\right)\Big]~.\label{s3massZ}
\end{align}
Here $\mathrm{Li}_2$ and $\mathrm{Li}_3$ are polylogarithms.  We plot the square of this result in figure \ref{s3massfig}.  We will make a few comments, and return to the bulk interpretation later:
\begin{figure}[t]
\centering
\includegraphics[height=8cm]{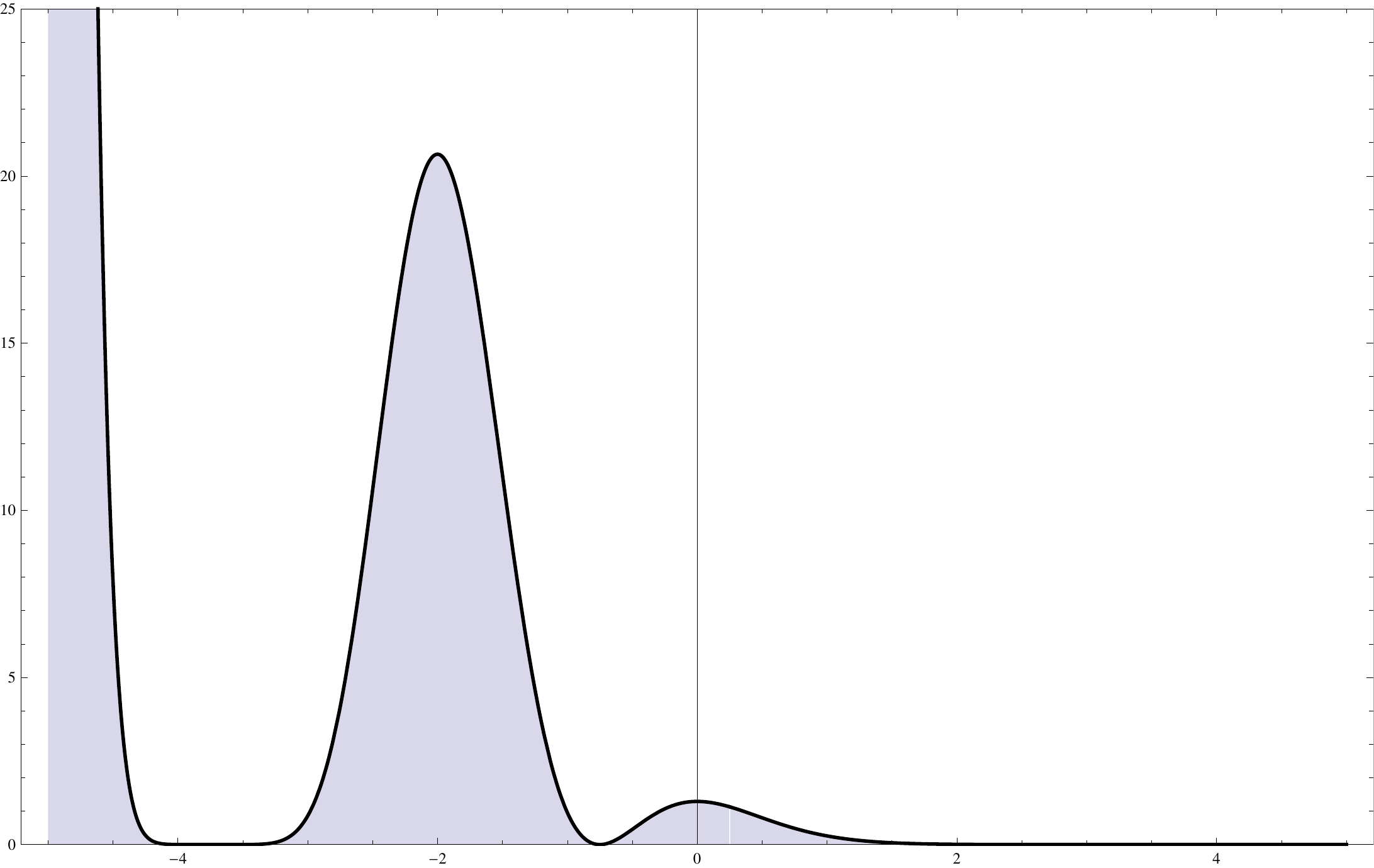} 
\caption{A plot of the probability distribution $|Z_{finite}|^2=|\Psi_{HH}|^2$ as a function of a constant mass perturbation $\sigma$ on the $S^3$, for $N=2$.}\label{s3massfig}
\end{figure} 
\begin{itemize}
\item There is a local maximum at $\sigma=0$, which is the conformally invariant point.  Zero-mode fluctuations are suppressed, which suggests that the dS solution of higher-spin gravity is perturbatively stable.  In fact since the two-point function of $J^{(0)}$ is negative-definite this perturbative stability extends to the other modes of $\sigma$ as well.  Using a concavity argument analogous to one we will outline in our discussion of the squashed sphere below, we can even prove that this maximum is a global maximum over all functions $\sigma(x)$ which are everywhere greater than $-\frac{3}{4}$.
\item The wave function squared vanishes for $\sigma=-\ell(\ell+2)-\frac{3}{4}$; this is because at these values there are fermion zero modes so the determinant vanishes.
\item At large constant positive $\sigma$ we have $Z\sim e^{-N\frac{\pi}{6}\sigma^{\frac{3}{2}}}$, so the probability distribution is exponentially suppressed.  We won't show it here, but this behavior is just that of the partition function on $\mathbb{R}^3$; this makes sense since when the mass is large the fluctuations should not be sensitive to the radius of curvature.  
\item At large constant negative $\sigma$ the wave function is divergent; the peaks between the zeros grow exponentially in $\sigma$ like $|Z|\sim e^{-\frac{N}{2}\sigma \log 2 }$.  This suggests that the wave function is non-normalizeable in the negative $\sigma$ direction, although we will not be able to decisively say this without a discussion of integration in other field directions that we elaborate on below.  We nonetheless see that the peak of the probability distribution is not at the dS invariant point.  We are inclined to interpret this as an instability of dS space in higher-spin gravity; the equilibrium field distribution at late times is far from its dS value.  We discuss this point further in section \ref{secmore}.  It is worth mentioning that the growth does not start until $O(1)$ values of $\sigma$, which our boundary conditions \eqref{altBC} tell us are actually bulk field values that are $O(\sqrt{N})$.  Thus from a bulk point of view the instability is a finite $N$ effect.  For comparison we will show in section \ref{einstsec} below that there is no such divergence in the wave function for a free bulk scalar in de Sitter space.
\end{itemize}


\section{The squashed $S^3$ partition function of the $Sp(N)$ model}\label{sqs3sec}

In this section we will compute the partition function of the $Sp(N)$ model on a squashed $S^3$, with $\sigma=0$. This is a purely metrical deformation, the space remains topologically an $S^3$ and we can continue to neglect the Chern-Simons sector.\footnote{It is worth noting that asymptotically de Sitter solutions with squashed $S^3$ boundary metric, known as Taub-NUT de Sitter space, exist in Einstein gravity with a positive cosmological constant, see for example \cite{Clarkson:2003wa}. It would be interesting to explore the existence of such solutions in the bulk Vasiliev gravity.}

\subsection{Squashed $S^3$ geometry}

The metric of the squashed $S^3$ is given by an $S^1$ fibered over an $S^2$ base space:
\begin{equation}
ds^2 = \frac{r^2}{4} \left( d\theta^2 + \cos^2\theta d\phi^2 + \frac{1}{(1+\alpha)} \left( d\psi + \sin \theta d\phi \right)^2 \right)~,
\end{equation}
with $\psi \sim \psi + 4\pi$, $\theta \in \left[-\frac{\pi}{2},\frac{\pi}{2}\right]$ and $\phi \sim \phi+2\pi$.  The squashed $S^3$ is a homogenous yet anisotropic space. When $\alpha=0$ the geometry becomes the round metric on $S^3$ expressed as a Hopf fibration. The Ricci scalar of the squashed $S^3$ is given by
\begin{equation}
R = \frac{2(3+4\alpha)}{r^2 (1+\alpha)}~,
\end{equation}
and becomes negative for $\alpha \in (-1, -3/4)$.  The volume is
\be
V=\frac{2\pi^2r^3}{\sqrt{1+\alpha}}~.
\ee
In what follows we set $r=1$ unless otherwise specified.

The eigenvalues of the conformal Laplacian on the squashed sphere are given by \cite{Hu:1974fs}:
\begin{equation}\label{evals}
\lambda_{n,q} = \left[ n^2 + \alpha \left( n - 1 - 2q \right)^2 -  \frac{1}{4(1+\alpha)} \right]~, \quad q = 0,1,\ldots n-1~, n=1,2,\ldots~,
\end{equation} 
with multiplicity $n$. The values of $\alpha$ which have vanishing eigenvalues are all less than  $\alpha = -3/4$ and in fact accumulate rapidly as $\alpha \to -1$. Based on our calculation of the previous section we might expect wild oscillation for $\alpha \in (-1, -3/4)$.

The determinant of the conformal Laplacian for the squashed sphere has been studied using zeta function regularization in \cite{SISSA-73-86-A,hep-th/0503238}, but it will be convenient for us to use a more numerical approach.  

\subsection{A New Regulator}

We could attempt to study the squashed sphere determinant using a simple hard eigenmode cutoff as in equation \eqref{hardcutoffsum} above, but this method is too clumsy to practically deal with the less symmetric geometry.  We will instead use a heat-kernel type regulator \cite{Vassilevich:2003xt}, which for a general set of eigenvalues $\lambda_i$ defines
\be
2\log Z_{finite}=-2\hat{S}_{ct}-N\sum_i \int_\epsilon^\infty\frac{dt}{t}e^{-t\lambda_i}~.
\ee
The justification for this definition is that for $\lambda\epsilon\ll 1$ we have
\be
-\int_\epsilon^\infty\frac{dt}{t}e^{-t\lambda_n}=-\Gamma(0,\lambda_n \epsilon)=\log\left(\lambda_n \epsilon e^{\gamma}\right)+O(\lambda_n \epsilon),
\ee
while for $\lambda\epsilon\gg 1$ we have 
\be
-\Gamma(0,\lambda \epsilon)=-e^{-\lambda \epsilon}\left[\frac{1}{\lambda\epsilon}+O\left(\frac{1}{(\lambda\epsilon)^2}\right)\right]~.
\ee
Thus this proposal is equivalent to the determinant for modes whose energies are less than a ``soft'' cutoff $\epsilon^{-1/2}$, and it cuts off the sum exponentially above the cutoff.  The factor of $e^{\gamma}$ is a wave function renormalization, similar to $a$ in the previous section.  

\begin{figure}
\begin{center}
\includegraphics[height=8cm]{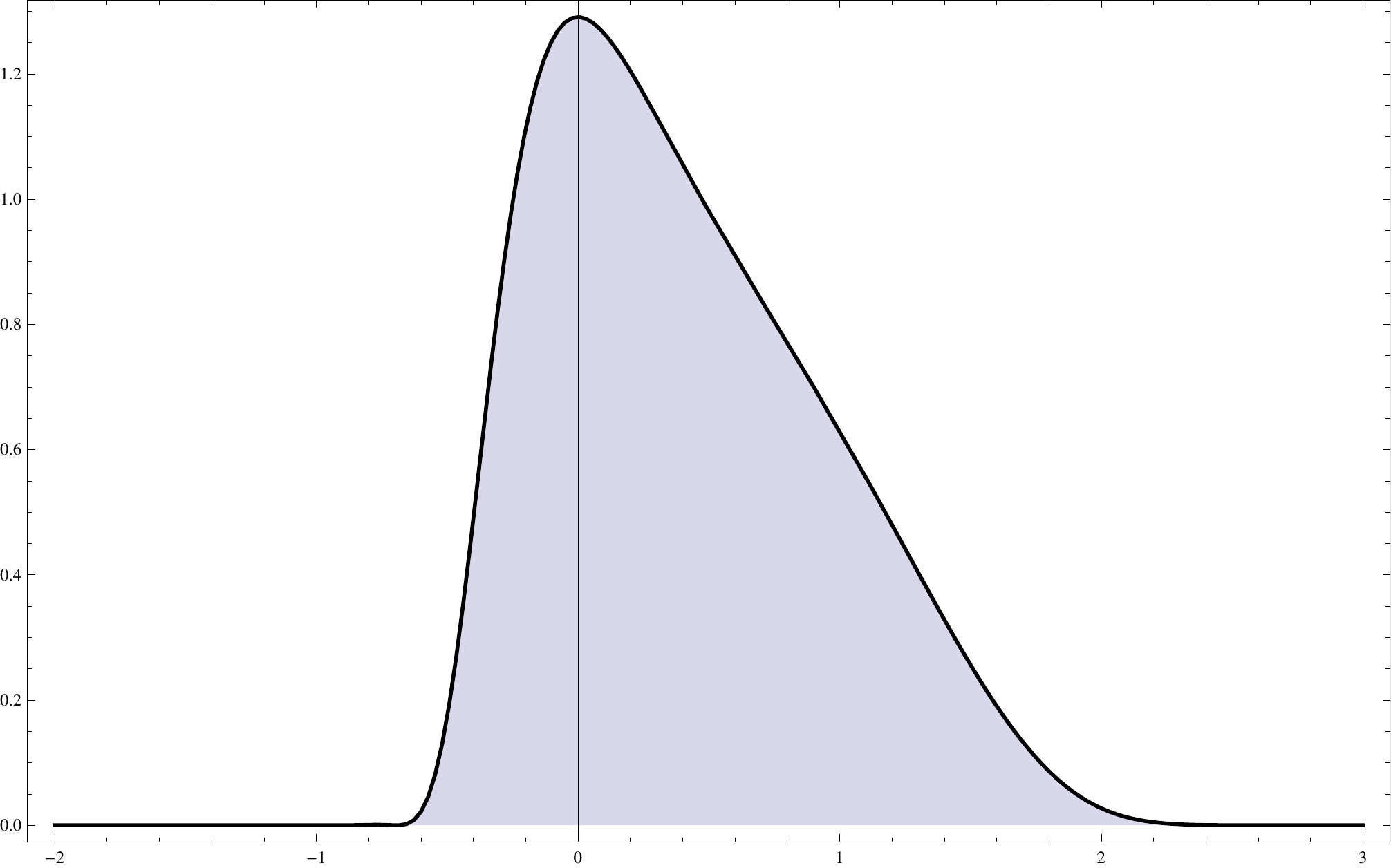}
\end{center}
\caption{\small{A plot of $|\Psi|^2$ with $N=2$ as a function of $\rho$, with $\rho$ related to the squashing $\alpha$ as $\alpha = e^{2\rho} - 1$. }}\label{sqs3fig}
\end{figure}

It is convenient to split the integral over $t$ into two pieces, which for the squashed sphere have the form
\be
2\log Z_{finite}= \text{det}_{UV} + \text{det}_{IR}~,
\ee
with 
\begin{align}
\text{det}_{UV}&\equiv -2\hat{S}_{ct}-N \int_\epsilon^\delta\frac{dt}{t}\sum_{n=0}^\infty \sum_{q=0}^{n-1}ne^{-t\lambda_{n,q}}\\
\text{det}_{IR}&\equiv -N \sum_{n=0}^\infty \sum_{q=0}^{n-1}n\int_\delta^\infty\frac{dt}{t}e^{-t\lambda_{n,q}}=-N \sum_{n=0}^\infty \sum_{q=0}^{n-1}n\Gamma(0,\lambda_{n,q}\delta)~.
\end{align}
Here $\delta$ is some small but $O(1)$ number, say $.01$.  The sum in $\text{det}_{IR}$ is easy to perform numerically, one needs to sum up to about $n=1/\delta$ to get good accuracy.  $\text{det}_{UV}$ requires a little more care, since the $\epsilon$ dependence needs to be treated analytically in order to extract the counterterms.  Our method is to apply the Euler-Maclaurin formula\footnote{In this formula $B_{2k}$ are the Bernoulli numbers, and the series on the right-hand side is asymptotic.  Truncating it at some finite $k$, the error is of order $\int_a^b dn |f^{(2k)}(n)|$.}
\be
\sum_{n=a}^{b} f(n)\sim \int_a^b dn f(n)+\frac{1}{2}\left(f(a)+f(b)\right)+\sum_{k=1}^{\infty} \frac{B_{2k}}{(2k)!}\left(f^{(2k-1)}(b)-f^{(2k-1)}(a)\right)
\ee
to each sum in $\text{det}_{UV}$, before doing the integral over $t$.  It is straightforward to see that the corrections to the formula are of increasing power in $t$, so if $\delta$ is small the integral is a good approximation to the sum.  In our calculations we included terms up to $k=2$ for both sums.  We can expand the result in powers of $t$ and then perform the integral; clearly only the lowest powers of $t$ can produce UV divergences so we can extract the divergences analytically and subtract them with an appropriate choice of $\hat{A}$ and $\hat{B}$.  Conveniently there is no quadratic divergence, so the wave function renormalization is apparently automatically handled by this regulator.  Finally we can take the limit $\epsilon\to0$ and add this analytic expression to our numerical result for $\text{det}_{IR}$ to get a plot of $|Z_{finite}|^2=|\Psi_{HH}|^2$ over a wide range of values of $\alpha$.  We show the result in figure \ref{sqs3fig}.

There are some things to notice about this picture:
\begin{itemize}
\item It is peaked at the round sphere $\alpha=0$.  Apparently the universe prefers being round to being squashed, at least when we condition all of the other sources to be zero.  The negative-definiteness of the stress tensor two-point function in fact shows that this peak is perturbatively stable under arbitrary metric deformations, with the only flat directions corresponding to three-dimensional diffeomorphisms and Weyl rescalings.\footnote{The simplest way to see this negative definiteness is to recall that this partition function is the inverse of the $O(N)$ model partition function, which has a reflection-positive path integral and a stress-tensor two-point function which is \textit{positive} definite.}  The peak gets sharper as $N$ becomes large.
\item It is exponentially suppressed both for large $\alpha$ and for $\alpha \to -1$; there does not appear to be any instability for $\alpha<-\frac{3}{4}$.  If one zooms in on this region oscillation is indeed seen as in the mass deformation case, but the peaks are exponentially damped rather than exponentially growing.
\item In fact we can argue that the round metric is a global maximum of the partition function over metrics with positive, but not necessarily constant, Ricci scalar.  The reason is that the negative definiteness of the stress tensor two-point function is true for any metric with such a Ricci scalar, so the partition function is real and weakly concave.  Any maximum must thus be a global maximum.\footnote{The condition on the Ricci scalar is necessary to ensure that in the $O(N)$ model the path integral can be done on a real integration contour, which is needed to conclude that its two-point function is positive.  It really is necessary, since for $\alpha<-\frac{3}{4}$ there are indeed other maxima and the concavity breaks down - in this case however these other maxima are only local and not global.}  This result is consistent with \cite{oki} from the mathematics literature.  Unfortunately we cannot decisively rule out other maxima with non-positive Ricci scalar which are higher than the round sphere. It would be very interesting to explore this possibility.

\end{itemize}


\section{The $S^2 \times S^1$ partition function of the `$U(-N)$' model}\label{s2s1sect}

We now turn to $S^2 \times S^1$.\footnote{We acknowledge here soon to appear related work \cite{alexalejandra}, where it is shown that in three-dimensional bulk de Sitter gravity there is a result similar to our main result of this section; the wave function with boundary geometry $T^2$ diverges at large conformal structure.}  Because of the topology the singlet constraint is now nontrivial; we need to deal with the Chern-Simons coupling.  The idea is that weakly gauging the $Sp(N)$ symmetry will dynamically enforce the singlet constraint \cite{arXiv:1109.3519}, and that in the limit $k\to\infty$ it will have no other affect on the dynamics.  We thus want to compute
\be
Z_{finite}\left[ S^2 \times S^1\right] =\lim_{k\to\infty}e^{-\hat{S}_{ct}}\int \mathcal{D}A\mathcal{D}\chi \; e^{-S_{CS}[A]-S_{bare}[\chi,A,g]}~.
\ee
$S_{bare}$ depends on the gauge field $A_i$ through covariant derivatives, while the Chern-Simons action $S_{CS}$ is independent of the metric:
\be
S_{CS}=\frac{k}{8\pi i}\int d^3 x \; \epsilon^{ijk} \mathrm{Tr}\left(A_i(\partial_j A_k-\partial_k A_j)+\frac{2}{3}A_i[A_j,A_k]\right)~.
\ee
  In the limit $k\to\infty$ we can evaluate the Chern-Simons path integral semiclassically by looking for flat connections.  On $S^3$ the only such connections are gauge equivalent to $A_i=0$, and the two sectors decouple.  On $S^2\times S^1$ there are flat connections which have nontrivial holonomy around $S^1$, and which have zero Chern-Simons action.  These connections do not decouple in the $k\to\infty$ limit and must be integrated over explicitly.  Our calculations in this section make extensive use of the tools developed in \cite{Brezin:1977sv,hep-th/9908001,hep-th/0310285,hep-th/0402219,157418}, but we will keep the discussion reasonably self-contained.

It is very convenient at this point to slightly modify the theory under discussion from an $Sp(N)$ theory of Grassmans to a $U(N)$ theory.\footnote{This theory is related to the bosonic $U(N)$ model by $N\to-N$ so it is sometimes whimsically called the $U(-N)$ model.  Group theoretically we have that $U(-N)=U(N)$ and $O(-N) = Sp(N)$ with symmetric and anti-symmetric representations switched \cite{Dunne:1988ih,Mkrtchyan:2010tt}.}  $\chi^a$ becomes a complex grassman field and instead of bilinears like $\chi^a \Omega_{ab}\chi^b$ the action and currents now have the form $\chi^\dagger \chi$.  It is very easy to go back and do all of our previous calculations again in this new model, the only difference is that the power of the determinant in equations \eqref{Zdet}, \eqref{Zdetfin} is $N$ instead of $N/2$.  No qualitative results are altered.  What this modification buys us is that we may now do a $U(N)$ gauge transformation to diagonalize the holonomy matrix for the Chern-Simons saddle point. Say we work on $S^2 \times S^1$ with coordinates such that the metric is
\be
ds^2=\beta^2 d\lambda^2+d\Omega_2^2~.
\ee
We will take $\lambda$ to have periodicity one, so the parameter $\beta$ measures the relative size of $S^1$ to $S^2$.  We will sometimes refer to its inverse as the temperature $T$, although the interpretation here has nothing to do with thermality.  In conventions where the covariant derivative is $D_i=\partial_i+A_i$, we may choose a gauge where any flat connection has the form
\be
A_i=
\begin{pmatrix}
-i \alpha_1 &0  &\ldots&0\\
0&-i\alpha_2 &\ldots&0\\
\vdots & \vdots & \ddots&\vdots\\
0 & 0 & \ldots & -i \alpha_N
\end{pmatrix}~.
\ee
This gauge fixing costs a Fadeev/Poppov determinant, and there is a remaining integral over the eigenvalues $\alpha_i$.  Integrating out the Grassman fields one finds \cite{arXiv:1109.3519,hep-th/0310285}:
\begin{align}\nonumber
Z_{finite}=\frac{1}{N!}\int \prod_n \left(d\alpha_n\right) &\exp \left[\sum_{n<m}\log \sin^2\left(\frac{\alpha_n-\alpha_m}{2}\right)-\hat{S}_{ct}\right]\\
&\times \prod_n \det\left(-(\partial_i+A_{i,n}) (\partial^i+A^i_n)+\frac{1}{4}\right)~.
\end{align}
To calculate the determinants one needs the eigenvalues\footnote{Here we have assumed periodic boundary conditions for the fermions.  Had we chosen antiperiodic boundary conditions we would have gotten the same result except with all $\alpha_i$ shifted by $\pi$.  This shift drops out of equation \eqref{alphaint} below since it can be absorbed into a change of variables in the integral over the $\alpha_i$.  The wave function is therefore independent of which sign choice we make for the fermion boundary conditions.}
\be
\lambda_{\ell,n,i}=\left(\ell+\frac{1}{2}\right)^2+\beta^{-2}(\alpha_i+2\pi i n), \qquad n\in \mathbb{Z}, \ell=0,1,2,\ldots.
\ee
The determinants can be regularized and renormalized using a hard cutoff as in equation \eqref{hardcutoffsum} above, the essential steps are described in section four of \cite{hep-th/0310285}.  The result is
\be\label{alphaint}
Z_{finite}=\frac{1}{N!}\int \prod_n \left(d\alpha_n\right) \exp \left[\sum_{n<m}\log \sin^2\left(\frac{\alpha_n-\alpha_m}{2}\right)-2\sum_{m=1}^\infty \frac{1}{m}z_S(e^{-\beta m})\sum_{n=1}^N\cos(m\alpha_n)\right]~,
\ee
where
\be
z_S(x)=x^{\frac{1}{2}}\frac{1+x}{(1-x)^2}~.
\ee
This differs from the bosonic $U(N)$ result of \cite{arXiv:1109.3519} only in the sign of the second term in the exponent.  

At finite $N$ it is difficult to learn more from this expression, but at large $N$ one can do the integral over the eigenvalues $\alpha_i$ using semiclassical techniques.  This was done in the AdS case by \cite{arXiv:1109.3519}, whose work we carry over almost verbatim.  There are two interesting regimes.  For temperatures that are $O(N^{0})$ the first term in the exponent is $O(N^2)$ while the second is only $O(N)$, so the saddle point is a small perturbation from what it would be with no matter present.  Since the first term is repulsive, the eigenvalues will be close to uniformly distributed.  Once the temperature becomes of order $\sqrt{N}$ however the second term also becomes $O(N^2)$.  This is because for large $T$ we have $z_S(e^{-\beta m}))\sim \frac{2T^2}{m^2}$.  The eigenvalue distribution then starts to be dominated by the second term, which is a potential that pushes the eigenvalues to $\pi$.  We now study this more quantitatively in both cases.

\subsection{Case 1: $T \ll \sqrt{N}$}

We  begin by considering `low' temperatures $T \ll \sqrt{N}$.  Since in this case the eigenvalues are roughly uniformly distributed it is convenient \cite{Brezin:1977sv} to introduce an eigenvalue distribution function $\rho(\alpha)$ which counts the density of eigenvalues around some particular value $\alpha$.  We then have:
\begin{equation}\label{densityaction}
\log Z_{finite} = N^2 \int d\alpha d\beta \rho (\alpha) \rho(\beta) \log \left| \sin \left( \frac{\alpha - \beta}{2} \right) \right| - 2  N \int d\alpha \rho(\alpha) \sum^\infty_{m=1} \frac{1}{m} z_S (x^m) \cos(m\alpha)~.
\end{equation}
The density of eigenvalues $\rho(\alpha)$ in the absence of the $U(N)$ vector matter fields is simply $\rho(\alpha) = \frac{1}{2\pi}$. This leads to the following $N^2$ contribution:
\begin{equation}
\log Z_{finite}[T] = - N^2 \log 2~.
\end{equation}
One can compute the $1/N$ subleading correction to the density, which we denote as $\tilde{\rho}/ N$, such that:
\begin{equation}
\rho(\alpha) = \frac{1}{2\pi} + \frac{1}{N} \tilde{\rho}(\alpha)~.
\end{equation}
Following the discussion of \cite{157418} we find the saddle point equation to be satisfied by $\tilde{\rho}(\alpha)$:
\begin{equation}
\mathbf{P} \int d\beta\tilde{\rho}(\beta) \cot \left( \frac{\alpha-\beta}{2}\right) = - 2 \sum_{m=1}^{\infty} z_S(x^m) \sin(m\alpha)~,
\end{equation}
which is solved by:
\begin{equation}\label{rhot}
\tilde{\rho} (\beta) = - \sum^\infty_{m=1} z_S (x^m) \frac{1}{\pi} \cos(m \beta)~.
\end{equation}
Here $\mathbf{P}$ means principal value.  Thus the partition function receives a correction given by:
\begin{eqnarray}
{\delta \log Z_{finite}[T]} &=&  \int d\alpha \tilde{\rho}(\alpha) \left( \int d\beta \tilde{\rho}(\beta) \log \left| \sin \left( \frac{\alpha - \beta}{2} \right) \right| - 2 \sum^\infty_{m=1} \frac{1}{m} z_S (x^m) \cos(m\alpha) \right)~,\\
\nonumber &=& \sum_{m,n}^\infty \int d\alpha d\beta \frac{\cos(n\alpha)}{\pi}\frac{\cos(m\beta)}{\pi} \log \left| \sin \left( \frac{\alpha - \beta}{2} \right) \right| z_S(x^m) z_S(x^n) + 2  \sum^\infty_{m=1} \frac{z^2_S (x^m)}{m}~.
\end{eqnarray}
This is the only piece of $Z_{finite}[T]$ that depends on temperature and thus the geometric data of $S^2\times S^1$. The first piece is somewhat subtle to evaluate. First we note that it vanishes for $m\neq n$. After some work we find:
\begin{equation}
\int_{-\pi}^{\pi} \int_{-\pi}^\pi d\alpha d\beta \cos(m\alpha) \cos(n\beta) \log \left| \sin \left( \frac{\alpha - \beta}{2} \right) \right| = -\frac{\pi^2}{m} \delta_{m,n}~. 
\end{equation}
Thus, putting everything together:\footnote{To obtain the result (\ref{lowtemp}) we have used that:
\begin{equation}
\int_{-\pi}^{\pi} \int^{\pi}_{-\pi} d\alpha d\beta \cos( m\beta) \log \left|  \sin \left( \frac{\alpha - \beta}{2} \right) \right| = 0~,
\end{equation}
for non-zero integer $m$.
}
\begin{equation}\label{lowtemp}
{\delta \log Z_{finite}[T]} =  \sum_{m=1}^\infty  \frac{1}{m} {z^2_S(x^m)}~.
\end{equation}
Due to a subtle cancelation in the signs, the above term in fact agrees with the answer for AdS$_4$ \cite{arXiv:1109.3519} in the low temperature regime. It is the partition function of a free gas of higher spin particles. 

\subsection{Case 2:  $T \gg \sqrt{N}$}

\begin{figure}[h]
\begin{center}
$\begin{array}{c@{\hspace{1in}}c}
\multicolumn{1}{l}{} &
	\multicolumn{1}{l}{} \\ 
\epsfxsize=2.8in
\epsffile{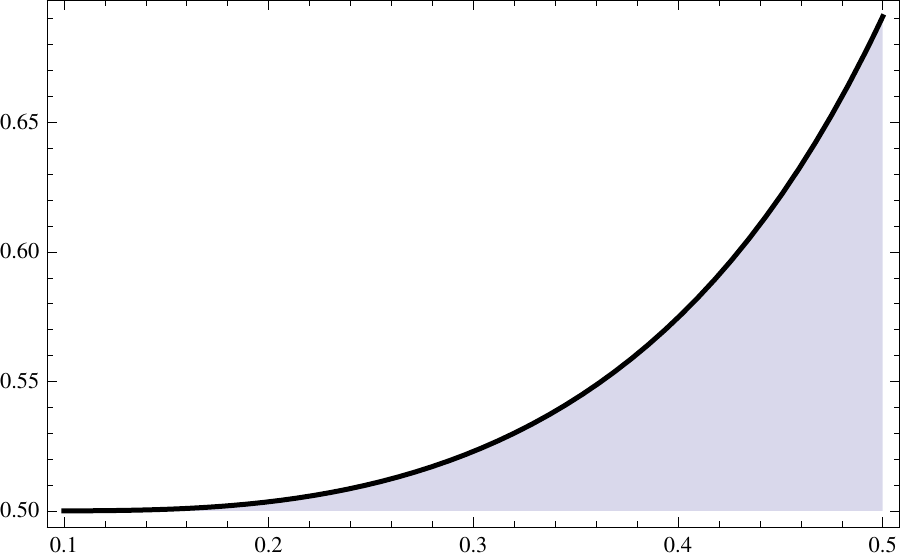}~ &
	\epsfxsize=3in
	\epsffile{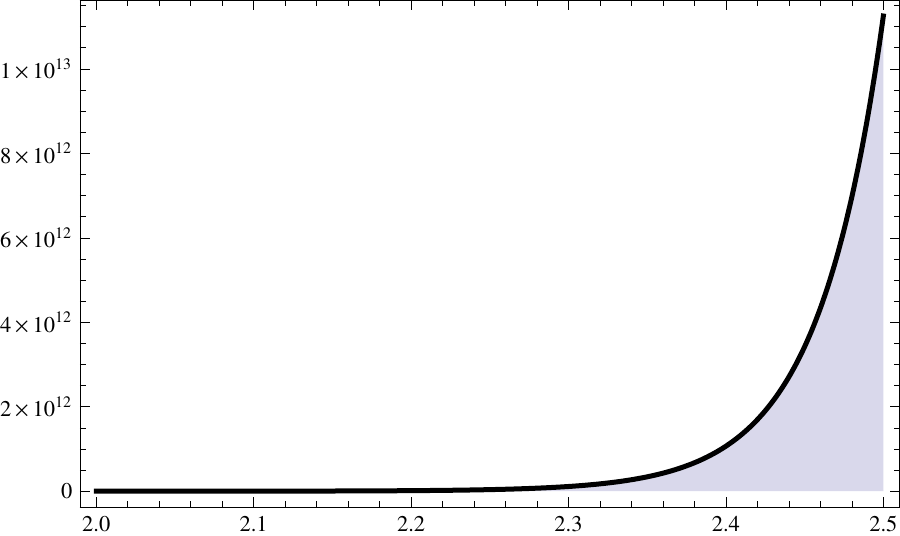} 
\end{array}$
\end{center}
\caption{\small{Left figure: $Z_{finite}[T]$ for $T < \sqrt{N}$. Right figure: $Z_{finite}[T]$ for  $T > \sqrt{N}$ with $N = 1$. For larger $N$ the curves become increasingly steep.}}
\label{s2s1}
\end{figure}

As mentioned above, the above analysis breaks down at sufficiently large temperatures since in that case the $\sim N^2$ and $\sim N$ terms begin to compete.  In the bosonic $U(N)$ model one finds a Gross-Witten \cite{157418} type transition for which the density of eigenvalues strongly favors that they all vanish. Interestingly, due to the sign flip in front of $N$ in (\ref{alphaint}) the potential at very large temperatures is $V(\alpha) \sim - \cos \alpha$ and forces the eigenvalues to be near $\alpha \sim \pi$.  We can then approximate \eqref{alphaint} by
\begin{equation}\label{s2s1largeT}
{ \log Z_{finite}[T]}= 3 \zeta(3) N T^2+\ldots~,
\end{equation}
where $\ldots$ are terms that are either subleading in $N$ or $1/T$ or are temperature independent.  As in the low temperature case, the result is similar to the bosonic $U(N)$ model studied in \cite{arXiv:1109.3519} in that the partition function {\it grows} with increasing $T$. In figure \ref{s2s1} we plot $\delta \log Z_{finite}[T]$ for small and large $T$.
We again make a few comments about this result:
\begin{itemize}
\item There does not seem to be even a local maximum for any finite value of $T$.  The probability distribution apparently prefers arbitrarily small radii for $S^1$.  This divergence is more subtle to interpret than that of the scalar; since the topology is not something that can be embedded into de Sitter space there is no particular reason to view it as an instability.  Rather it is a statement about a strange type of ``big bang'' where the universe is created with $S^1\times S^2$ topology.  Perhaps surprisingly, we will show in section \ref{einstsec} that it is actually possible to reproduce this type of divergence in standard Einstein gravity with positive $\Lambda_{bulk}$.  
\item The divergence depends crucially on the topology - the squashed sphere also locally has the isometry of $S^1 \times S^2$ but small $S^1$, which corresponded to $\alpha\to -1$, was exponentially suppressed.
\item In the bosonic $U(N)$ model the result at large temperature is \cite{arXiv:1109.3519} $\log Z_{finite}=4\zeta(3)N T^2$.  This is clearly not related to \eqref{s2s1largeT} by analytic continuation $N\to-N$; the reason is that the eigenvalues cluster at $\pi$ instead of $0$.  We view this as a nice demonstration that dS and AdS are not in general related by analytic continuation at the nonperturbative level, even though they are in perturbation theory in many cases \cite{Maldacena:2002vr,Harlow:2011ke,McFadden:2009fg}.
\end{itemize}

\section{Comparison with Einstein Gravity}\label{einstsec}

In this section we review the Hartle-Hawking wave function for a free massive scalar in dS$_4$ with boundary geometry $S^3$ and for gravity with boundary geometry $S^2\times S^1$.  
\subsection{Free Scalar}
The wave function of a free massive scalar with future geometry $S^3$ can be lifted from the AdS calculation done in appendix C of \cite{Harlow:2011ke}, along with the analytic continuations $\chi=\tau-\frac{i \pi}{2}$ and $\ell_{ads}=i \ell_{ds}$. This is in coordinates with metric
\be
ds^2=-d\tau^2+\cosh^2\tau d\Omega_{3}^2~,
\ee 
where we have set $\ell_{dS} = 1$.
Specializing the general result of \cite{Harlow:2011ke} to $d=3$ and $\tilde{\Delta}=2$, and also assuming that only the zero mode is turned on, we have
\be
\log \Psi[\tau,\phi]=i\pi^2\cosh^3\tau \phi_0^2\left[-1+i(\sinh\tau)^{-1}+\ldots\right]~.
\ee
To compare to a CFT calculation we define $\tilde{\sigma}_0=\phi \cosh\tau$, after which we see that
\be
|\Psi|^2=e^{-\pi^2\tilde{\sigma}^2}~.
\ee
Changing to the $\sigma$ basis we have
\be
|\Psi|^2=e^{-\frac{\pi^2\sigma^2}{8}}~.
\ee
The points to notice are that the divergent counterterm is pure phase and the wave function is square-normalizeable.  There is no sign of the divergence we found in the $Sp(N)$ model as a function of the mass deformation.

\subsection{Einstein Gravity with Positive Cosmological Constant and ``Thermal'' Boundary Conditions}

In this section we present a $\Lambda_{bulk} > 0$ version of the semiclassical Einstein gravity calculation \cite{Hawking:1982dh,Witten:1998zw} of the wave function with $S^{d-1}\times S^1$ boundary conditions  We will specialize to $d=3$ at the end.  We are interested in compact complex solutions of the Lorentzian action
\be
S_L=\frac{1}{16\pi G}\left[\int d^{d+1}x\sqrt{-g}\left(R-d(d-1)\right)+2\int d^d x \sqrt{\gamma}K\right]~,
\ee
which have a single boundary with topology $S^{d-1}\times S^1$ and induced boundary metric
\be
\gamma_{ij}dx^i dx^j=r_c^2\left(\left(\frac{\beta d\theta}{2\pi}\right)^2+d\Omega_{d-1}^2\right)~.
\ee
We have chosen the cosmological constant here so that the dS radius is one.  The action evaluated on such a solution is
\be
S_L=\frac{1}{8\pi G}\left[d\int d^{d+1}x\sqrt{-g}+2\int d^d x \sqrt{\gamma}K\right]~.
\ee
The only solutions with $SO(d)\times SO(2)$ symmetry are analytic continuations of dS/Schwarzchild:
\be
ds^2=-\frac{dr^2}{f(r)}+f(r)d\lambda^2+r^2 d\Omega_{d-2}^2~,
\ee
with 
\be
f(r)=r^2-1+\alpha r^{2-d}~.
\ee
The coordinate $r$ runs over a range $r\in(r_0,r_c)$, where for there to be no boundary at $r_0$ we must choose $r_0$ to be a root of $f(r)$.  The coordinate $\lambda$ we take to be periodic with period $\lambda_0$.  The trace of the extrinsic curvature at the boundary is
\be
K=-\sqrt{f}\left(\frac{f'}{2f}+\frac{d-1}{r}\right)|_{r=r_c}~,
\ee
and the action evaluates to 
\be
iS_L=\frac{i \Omega_{d-1}\lambda_0}{8\pi G}\left[-(d-1)r_c^{d-2}(rc^2-1)-\frac{d}{2}r_0^{d-2}+\frac{d-2}{2}r_0^d\right]~.
\ee
To match the periodicity of the $S^1$ at the boundary, we must have
\be\label{lambbeta}
\lambda_0=\frac{r_c\beta}{\sqrt{f(r_c)}}=\frac{\beta}{\sqrt{1-\frac{1}{r_c^2}}}\left(1-\frac{\alpha}{2r_c^d}+\ldots\right)~,
\ee
where $\ldots$ indicates terms that fall off faster as $r_c\to\infty$.  Inserting this into the action, we find
\be\label{s2s1bulkact}
iS_L=\frac{i \Omega_{d-1}\beta}{8\pi G}\frac{1}{\sqrt{1-\frac{1}{r_c^2}}}\left[(d-1)r_c^{d-2}(r_c^2-1)+\frac{1}{2}r_0^{d-2}(1+r_0^2)+O(1/r_c)\right]~.
\ee
Note that all divergent terms as $r_c\to \infty$ are pure phase, as argued for generally above equation \eqref{Zdetfin}.  In odd dimensions the square root in the denominator just shifts around the coefficients of the divergences, while in even dimensions it also contributes a phase to the finite piece.  

Finally we observe that $r_0$ and $\lambda_0$ are related by the requirement that the (complex) geometry be regular at $r_0$.  There are two ways for this to happen.  The first is if $r_0$=0; in this case $\alpha=0$ and the geometry closes off because the $S^{d-1}$ shrinks to zero size.  We may then set $\lambda_0$ at will to match $\beta$ as in equation \eqref{lambbeta}.  This geometry is a quotient of pure de Sitter space, and its wave function $\Psi=e^{iS_L}$ is pure phase.  The other option is for the $S^1$ to shrink to zero size at finite radius of the $S^{d-1}$, but in a way that the geometry caps smoothly.  This requires
\be
\lambda_0=\pm\frac{4\pi i r_0}{d-2-dr_0^2}~,
\ee
or equivalently 
\be
r_0=\pm \frac{2\pi i}{d\beta}\left(-1\pm\sqrt{1-\frac{d(d-2)\beta^2}{4\pi^2}}\right)+O(1/r_c)~.
\ee
Note that there are four possible complex solutions.  Intuitively the relative sign arises from whether $r_0$ is placed at the black hole or cosmological horizon of the continued de Sitter/Schwarzchild, while the overall sign is a choice of branch in how we continue the regularity condition away from real Euclidean geometries.  Without some sort of non-perturbative information about the Einstein gravity path integral we have no way of deciding how many of these solutions should contribute to the semiclassical limit of the wave function.  The most conservative thing to do is to acknowledge that any of them give a valid solution of the Wheeler deWitt equation, and study each in turn.

We are especially interested in the limits of high and low temperature.  For low temperature we have $r_0\to \pm \sqrt{\frac{d-2}{d}}$, which interestingly is the Nariai value of $r_0$.  The finite part of the action at low temperature is then
\be
iS_{finite}=\frac{i\Omega_{d-1}\beta}{8\pi G}(\pm)^{d-2}\frac{d-1}{d}\left(\frac{d-2}{d}\right)^{\frac{d-2}{2}}~.
\ee
This is a pure phase in the wave function that oscillates as $\beta\to\infty$.  The behavior at high temperature is more interesting, there are two types of solution: $r_0\to\pm\frac{i(d-2)\beta}{4\pi}$ or $r_0\to \pm \frac{4\pi i}{d\beta}$.  In the first type the action vanishes as $\beta\to 0$.  In the second type the action diverges, as a phase in the wave function for even dimension and as a diverging real part in odd dimension.\footnote{The divergence found by Castro and Maloney \cite{alexalejandra} for $d=2$ does not contradict this result because the solutions we are considering are only valid for $d\geq 3$.}  In particular for $d=3$, the behavior is
\be
iS_{finite}=\pm\frac{16\pi^3}{27G \beta^2}~.
\ee
For the upper choice of sign this is exactly the scaling with $G$ and $\beta$ that we found in the fermionic $U(N)$ model in equation \eqref{s2s1largeT}!  Apparently the divergence of the wave function at large temperature in the dual field theory is capturing something about de Sitter space that is present even in Einstein gravity.  It is interesting to note that the saddle points which control this limit are dS/Schwarzchild black holes with large and purely imaginary mass.  

\section{More on the Scalar Divergence}\label{secmore}

In this section we return to the divergence at large negative $\sigma_0$ we found in section \ref{sig0sec}.  We first discuss the extent to which it can be interpreted as a non-normalizeability of the wave function.  We then study the partition functions of the critical $O(N)$ and $Sp(N)$ theories on $S^3$ as a function of a constant mass deformation.  We'll see that in the $Sp(N)$ case it is actually impossible to define the critical theory on $S^3$ nonperturbatively in $N$.  Finally we will return to the question of the bulk interpretation of the scalar sector, concluding that the most reasonable interpretation is that it indicates a non-perturbative instability of Vasiliev's theory in de Sitter space.    

\subsection{Normalizeability?}\label{normal}

Because we are computing conditional probabilities where the sources for higher-spin fields are set to zero, it is possible that the exponential divergence we found in section \ref{sig0sec} does not actually indicate non-normalizeability of the wave function.  For example the function $\frac{1}{\sqrt{x^2+y^2}}$ is integrable near $x=y=0$, but if we set $y=0$ and try to perform the $x$ integral we will find a spurious divergence.  We can fix this by setting $x$ to be non-zero and then integrating over $y$ in some finite region around $y=0$.  The resulting distribution for $x$ will then be integrable.  In this subsection we will suggest an argument that integrating over any finite number of the zero modes of the higher spin sources will not resolve the apparent non-normalizeability of the wave function in the scalar zero mode, but to make this argument precise the higher-spin sources need to be studied in much more detail than we provide here.  

In the presence of a constant mass deformation the higher-spin currents can be modified in such a way that they are still covariantly conserved.  For example the stress tensor \eqref{stresstensor} picks up an extra term $\frac{\Omega_{ab}}{4}\sigma \chi^a\chi^b$.  More generally the spin $s$ current has the heuristic form
\be
J^{(s)}_{\phantom{(s)}\mu_1\ldots\mu_s}=\Omega_{ab}\left(\chi^a\partial_{\mu_1}\ldots \partial_{\mu_s}\chi^b+\ldots +g_{\mu_1\mu_2}\ldots g_{\mu_{s-1}\mu_s}\sigma^{s/2}\chi^a \chi^b\right)~,
\ee
The indices are symmetrized and have traces subtracted in the appropriate way.  We can turn on linearized sources for these operators in the action
\be
\delta S_{CFT}=\sum_s \int d^3 x \sqrt{g}h^{\mu_1\ldots \mu_s} J^{(s)}_{\phantom{(s)}\mu_1\ldots\mu_s}~.
\ee
To Gaussian order the source dependent piece of the partition function is
\be
\log Z_{finite}\supset\frac{1}{2}\int d^3x \sqrt{g}(x)d^3y\sqrt{g}(y)\langle J^{(s)}_{\phantom{(s)}\nu_1\ldots\nu_s}(x)J^{(s)}_{\phantom{(s)}\nu_1\ldots\nu_s}(y)\rangle h^{\nu_1\ldots\nu_s}(x)h^{\nu_1\ldots\nu_s}(y)~.
\ee
At finite $\sigma$ the two-point function of $J^{(s)}$ in this expression is calculated using massive propagators.  Various quadratic counterterms in $h_{\mu_1\ldots\mu_s}$ are needed to to make the correlator finite which we will not be explicit about.  Evaluating this integral on $S^3$ is difficult, it requires higher-spin spherical harmonics.  On $\mathbb{R}^3$ however it is easy to see that the integral is proportional to a power of $\sigma$.  Performing the Gaussian integral over $h_{\mu_1\ldots \mu_s}$ will bring down the square root of the inverse of this power of $\sigma$ into the integration measure for $\sigma$, but it will never be able to beat an exponential divergence of the type found in section \ref{sig0sec}.  Unfortunately although the answer on $\mathbb{R}^3$ is a good approximation to the $S^3$ calculation in most of the $\sigma$ plane at large $\sigma$, it fails precisely on the negative real $\sigma$ axis where the divergence happens.  The full $S^3$ calculation is beyond the scope of this work, but we note that generically one encounters powers of $\sigma$ times sums of the form
\be
\sum_{\ell=0}^\infty (\ell+1)^2 \frac{1}{(\ell(\ell+2)+\frac{3}{4}+\sigma)^2}.
\ee
This sum can be performed analytically, and its inverse does not approach zero at large negative $\sigma$.  Instead it oscillates with constant amplitude.  To the extent that this sum is a good model for the full set of terms that appear we see that integrating over $h_{\mu_1\ldots \mu_s}$ will not affect the non-normalizeability of the integral over $\sigma$.

\subsection{The Partition Function of the critical $O(N)$ Theory on the Three Sphere}

As explained in appendix \ref{altquantapp}, one expects the wave function in the field basis to be computed by the interacting critical $Sp(N)$ theory reached in the IR of a ``double trace'' flow from the free $Sp(N)$ theory.  This theory ends up being subtly inconsistent, so as a warmup in this section we compute the partition function on $S^3$ as a function of a constant mass deformation of the \textit{bosonic} critical $O(N)$ model conjectured to be dual to higher spin gravity in AdS$_4$.  The tools we develop for this more familiar case we will then apply to the $Sp(N)$ theory in the following subsection.

\begin{figure}[t]
\centering
\includegraphics[height=6cm]{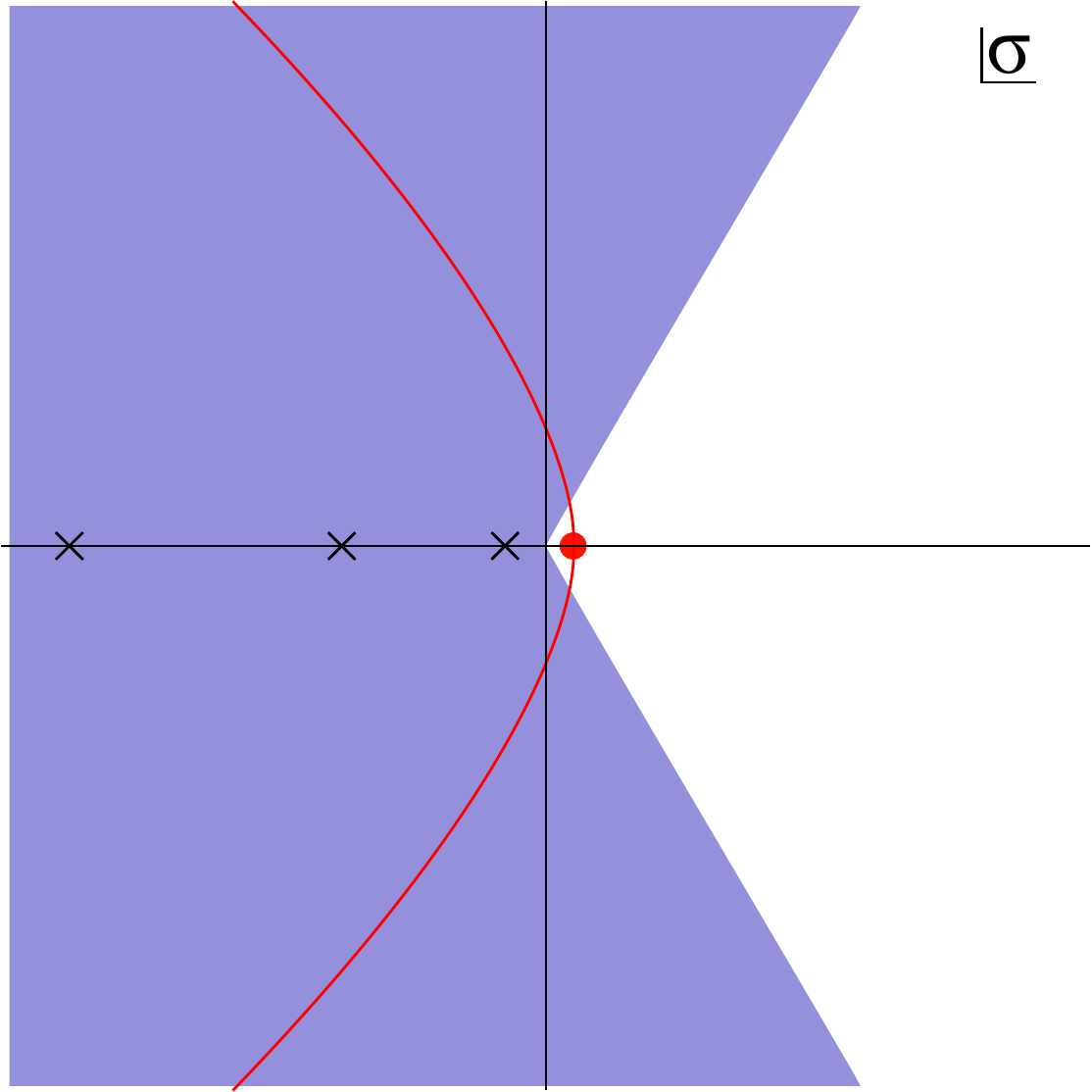} \qquad \includegraphics[height=5cm]{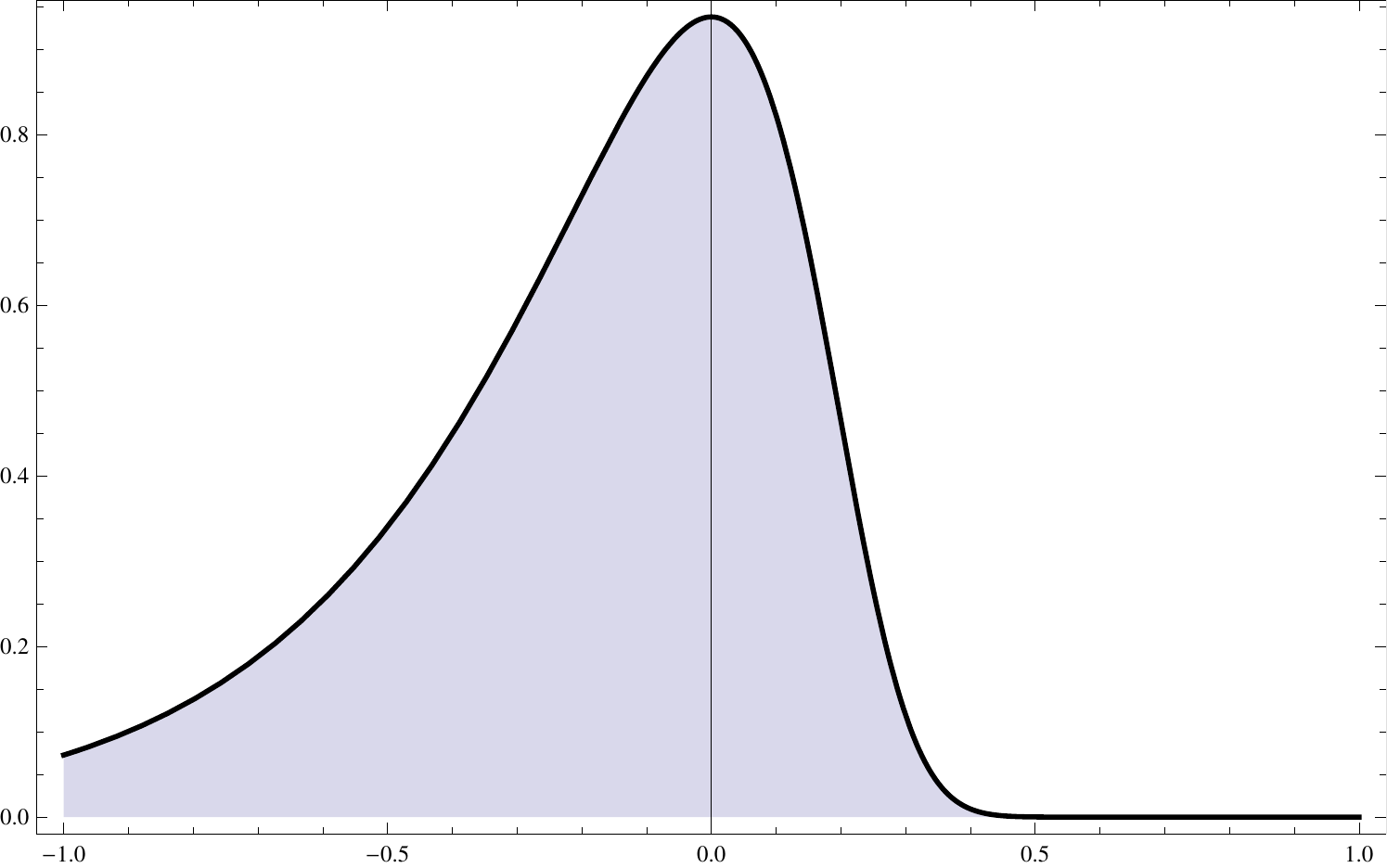}
\caption{Important features of the $\sigma$ plane, and a plot of our large $N$ result for $\exp\left[\frac{1}{N} \log Z_{crit}[\tilde{\sigma}]-\frac{f  \pi^2\tilde{\sigma}^2}{16}\right]$ as a function of $\tilde{\sigma}$.}\label{critadsplot}
\end{figure}

We first define\footnote{In this subsection only $\phi$ is a boundary field, not a bulk field.}
\be
Z_{free}[\sigma]\equiv e^{ -N I[\sigma]}=\int \mathcal{D}\phi \exp \left[-\frac{1}{2}\int d\Omega_3 \left\{(\partial\phi)^2+\frac{3}{4}\phi^2+\sigma \phi^2\right\}\right]
\ee
and
\be
Z_{crit}[\tilde{\sigma}]=\int \mathcal{D}\phi \exp \left[-\frac{1}{2}\int d\Omega_3 \left\{(\partial\phi)^2+\frac{3}{4}\phi^2+\frac{f}{4}\tilde{\sigma} \phi^2+\frac{f}{4N}(\phi^2)^2\right\}\right]~.
\ee
The $\phi$'s have vector indices running from $1$ to $N$ that are contracted in the obvious way and $f \in \mathbb{C}$ is some constant.  The function $I(\sigma)$ is defined as the right hand side of equation \eqref{s3massZ} divided by $N$.  Note the free $O(N)$ partition function is the inverse of the free $Sp(N)$ partition function.  As discussed in more detail in the appendix the critical partition function is a wave function (here a radial one since we are in AdS) in the field basis.  These partition functions are related by
\be\label{sigmaNint}
Z_{crit}[\tilde{\sigma}]=\sqrt{\det\left(\frac{-N}{2\pi f}\right)}e^{\frac{fN}{32}\int d\Omega_3\tilde{\sigma}^2}\int \mathcal{D}\sigma \exp \left[N\int d\Omega_3\left(\frac{\sigma^2}{2f}-\frac{1}{4}\sigma\tilde{\sigma}-\frac{1}{2\pi^2}I[\sigma]\right)\right]~,
\ee
which is a special case of \eqref{alt2st} in the appendix.  At large $N$ this integral can be evaluated semiclassically by looking for a saddle point $\sigma_c$ of the integral over $\sigma$.  When $\tilde{\sigma}$ is constant, the saddle point equation is be algebraic (see for example the appendix of \cite{arXiv:1105.4598}):
\be
4\pi \tilde{\sigma}_0=\sqrt{1-4\sigma_c}\cot \left(\frac{\pi}{2}\sqrt{1-4\sigma_c}\right)~.
\ee
For real $\tilde{\sigma}_0$ this equation has infinitely many solutions, one greater than $-3/4$ and the rest less than it.  The one which is greater than $-3/4$ approaches $-3/4$ as $\tilde{\sigma}_0\to-\infty$ and approaches $+\infty$ as $\tilde{\sigma}_0\to+\infty$.  To determine which is the relevant one to use in the semiclassical approximation we need to study the integration contour for $\sigma$.  For $f$ real and positive we must take the $\sigma$ contour to come in from a wedge of angle $\pi/2$ centered on the negative imaginary axis and go out through a wedge of angle $\pi/2$ centered on the positive imaginary axis.  For example we can just take it to run along the imaginary axis.  As $f$ becomes large however the behavior of the integrand is controlled by $I(\sigma)$ for a very long time before the term $\frac{\sigma^2}{2f}$ takes over.  For the integral to have a limit as $f\to\infty$ it is necessary that we can deform the original integration contour smoothly into one that remains convergent as $f\to\infty$.\footnote{The partition function will still be divergent in this limit because of the local term in front of the integral in equation \eqref{sigmaNint}, but we want to make sure that all divergence has been isolated in this term.}  The asymptotic behavior for $I(\sigma)$ away from the negative real axis is
\be\label{largeI}
I(\sigma)\sim -\frac{\pi}{6}\sigma^{\frac{3}{2}}~,
\ee
so in order for the the limit $f\to\infty$ to exist the defining contour must be deformable to one that enters from the region $-\pi<\theta<-\pi/3$ of the sigma plane and exits through the region $\pi/3<\theta<\pi$.  We shade these regions in blue in figure \ref{critadsplot}, which illustrates the relevant features of the $\sigma$ plane.  
The singularities of the integrand, which are the values of $\sigma$ where the determinant vanishes, are marked as $x$'s. We are not allowed to deform the integration contour across them.  The saddle point with $\sigma_c>-3/4$ is marked in red for a typical value of $\tilde{\sigma}_0$.  The lesson however is that it is clear from the figure that the defining contour can always be deformed into the red stationary phase contour that passes through the red saddle point, so we can approximate $Z_{crit}$ at large $N$ as the contribution of this saddle point only.  We plot our result for the finite part of $Z_{crit}$ in figure \ref{critadsplot}.  For generic $\tilde{\sigma_0}$ the saddle point equation must be solved numerically, but in asymptotic limits it is not hard to see that
\begin{align}
\frac{1}{N}\log Z_{crit}[\tilde{\sigma}]-\frac{f\pi^2}{16}\tilde{\sigma}^2&\approx -\frac{2\pi^4}{3}\tilde{\sigma}^3  \qquad\qquad \tilde{\sigma}\to\infty\\
\frac{1}{N}\log Z_{crit}[\tilde{\sigma}]-\frac{f\pi^2}{16}\tilde{\sigma}^2&\approx\frac{1}{16}\left(\log 4-\frac{3\zeta(3)}{\pi^2}\right)-\pi^2 \tilde{\sigma}^2\qquad\qquad \tilde{\sigma}\to 0\\
\frac{1}{N}\log Z_{crit}[\tilde{\sigma}]-\frac{f\pi^2}{16}\tilde{\sigma}^2&\approx\frac{3\pi^2}{8}\tilde{\sigma}-\frac{1}{2}\log \left(-\frac{1}{\pi^2\tilde{\sigma}}\right)\qquad\qquad \tilde{\sigma}\to -\infty~.
\end{align}

\begin{figure}[h!]
\centering
\includegraphics[height=1.5cm]{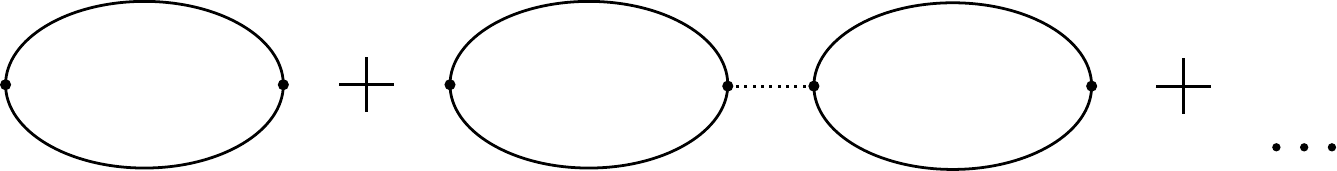}
\caption{The first few diagrams that contribute at order $N$ to the $\phi^2$ two point function.}\label{critdiags}
\end{figure}

Finally we observe that when $\tilde{\sigma}=0$, the relevant saddle point is $\sigma_c=0$.  This is consistent with the perturbation theory usually used to study correlation functions in the critical $O(N)$ model; recall that in this perturbation theory at large $N$ one sums over ``cactus'' diagrams.  For example the two-point function of $\phi^2$ at large $N$ is dominated by a set of diagrams shown in figure \ref{critdiags}.
Here the dashed line is a $\sigma$ propagator $-\frac{N}{f}$.  These diagrams are a geometric series which can be summed to see that $\phi^2$ has dimension two when the momentum transfer is low compared to $f$.  Using these diagrammatics we can confirm that the one point function of $\sigma$ is indeed zero as long as we normal order $\phi^2$ in the action.

\subsection{The Partition Function of the Critical $Sp(N)$ Theory on the Three Sphere}

\begin{figure}[t]
\centering
\includegraphics[height=6cm]{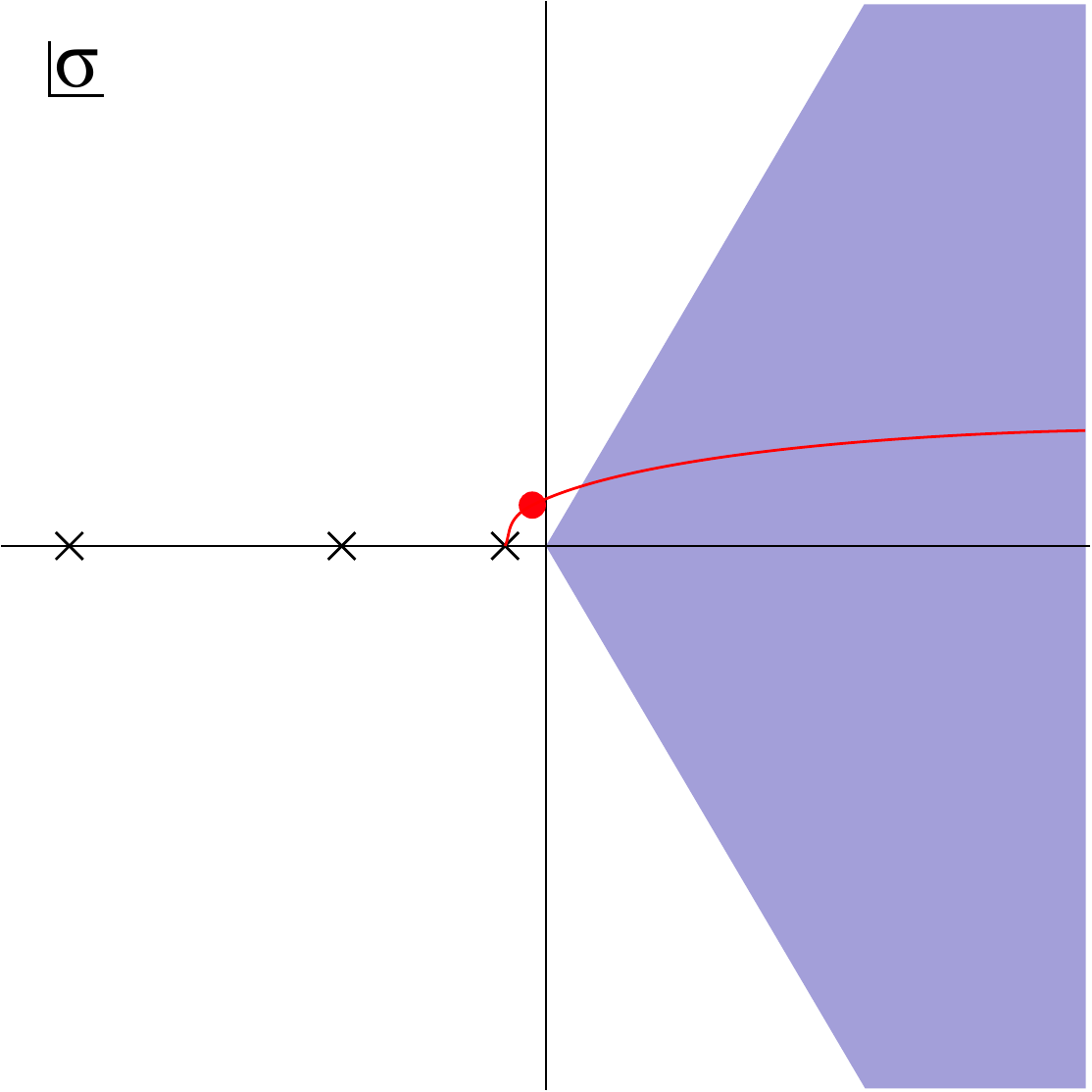}
\caption{Important features of the $\sigma$ plane for the dS case.}\label{critds}
\end{figure}

We now attempt to study the critical $Sp(N)$ theory using the same tools just applied to the $O(N)$ model.  As we explain in detail in appendix A, we would like to interpret this partition function as the field basis representation of the Hartle-Hawking wave function.  From equation \eqref{alt2st} we are interested in the integral 
\be\label{sigmaNintSp}
Z_{crit}[\tilde{\sigma}]=\sqrt{\det\left(\frac{-N}{2\pi f}\right)}e^{-\frac{iN}{2\epsilon}\int d\Omega_3\tilde{\sigma}^2}\int \mathcal{D}\sigma \exp \left[N\int d\Omega_3\left(\frac{\sigma^2}{2f}+\frac{i}{4}\sigma\tilde{\sigma}+\frac{1}{2\pi^2}I[\sigma]\right)\right]~,
\ee
where we have used the results of section A.3.  Here $\epsilon$ is a real and positive length cutoff defined so that $f=f_0\epsilon^{-1}$.  We will get to the critical theory by taking $\epsilon\to0$ with $f_0$ fixed.  The phase of $f_0$ is a free choice, we will consider various possibilities.  The saddle point equation is
\be
4\pi i \tilde{\sigma}_0=\sqrt{1-4\sigma_c}\cot \left(\frac{\pi}{2}\sqrt{1-4\sigma_c}\right)~,
\ee
so there is still a saddle point $\sigma_c$ which is zero when $\tilde{\sigma}=0$.  As in the $O(N)$ model the ``cactus'' perturbation theory is based on this saddle point.  There are again however lots of other saddle points; essentially there is one associated with each of the zeros of the functional determinant of the conformal Laplacian.  In figure \ref{critds} we show the important features of the $\sigma$ plane analogous to those in the $O(N)$ model.

There is a clear difference in that the allowed region in which the integration contour may approach infinity, again shaded blue, is much smaller.  Unlike the AdS case, there are no singularities in the blue region.  This means that the integral on any contour that starts and ends at infinity in the blue region must be zero!  The only way to get a nontrivial result on a contour that runs from infinity to infinity is for the contour to come from and/or go to infinity along the negative real axis, where the approximation \eqref{largeI} breaks down.  For example if $f_0$ is real and negative then we can take the defining contour to run along the real axis.  The term $\frac{\sigma^2}{2f}$ in the exponent of the integrand will render the integral convergent, but the integral will be dominated by the region of very negative $\sigma$.  The limit $f\to\infty$ will not exist.  It is not difficult to see that there is no choice of phase for $f_0$ where the defining contour can be deformed in such a way to produce a good limit as $f\to\infty$.  In particular there is no defining contour from infinity to infinity on which the integral can be semiclassically approximated by the cactus perturbation theory; the stationary phase contour which passes through the ``cactus'' saddle point runs from $-3/4$ to infinity and is indicated in figure \ref{critds}.  Thus we see that non-perturbatively in $N$ the critical $Sp(N)$ theory on $S^3$ does not exist.\footnote{One might be tempted to take the red contour in figure \ref{critds} as the \textit{defining} contour for the critical $Sp(N)$ theory.  This would be giving up on the double trace flow interpretation, and it would be unpleasantly ambiguous.  Why should the path integral stop at $\sigma=-3/4$ and not some other value?  We could try to make it sound less ambiguous by defining the end point as the least negative eigenvalue of the conformal Laplacian, but the path integral would then depend non-locally on the metric and calling it a field theory would be questionable.  This type of metric dependence does not seem consistent with what we would expect from the bulk.  We do not find options of this kind appealing.}

\subsection{Bulk Interpretation}

What are we to make of the scalar divergence and the non-existence of the critical $Sp(N)$ theory at finite source?  That both of these theories should have well-defined partition functions with finite sources turned on is equivalent to the statement that in the bulk there are limiting probability distributions for $\sigma$ and $\tilde{\sigma}$ which are well defined at late times.  This is essentially the assumption that the late time behavior is asymptotically de Sitter in the Fefferman-Graham sense.  We have uncovered a contradiction in this assumption for the higher-spin theory dual to the free $Sp(N)$ model.  The only way out is to conclude that this model does not in fact describe a theory which is asymptotically de Sitter at late times.  What instead must happen is some sort of instability that allows $\sigma$ to find its way out to large negative values.  Recall that in quantum mechanics non-normalizeability of the ground state wave function is the key indicator for an unstable system.  For example a particle in an exponential potential has a ground state wave function which grows linearly with distance on the decreasing side of the exponential.  Since the Hartle-Hawking wave function is in some sense supposed to be describing the equilibrium configuration of the theory, we conclude that in this equilibrium the field $\sigma$ tends to be far from its de Sitter invariant value.  Of course since we were not able to decisively prove non-normalizeability for the technical reasons described in section \ref{normal}, we can't completely rule out the possibility that once all of the sources for higher spin fields are integrated over there is a miraculous pressure to positive $\sigma$ that restores the de Sitter invariance.  We have found no indication so far that this happens, and we find it simpler to envision that higher spin gravity in de Sitter space is unstable.

\section{Three-manifolds and the `big bang'?}\label{bigbang}

Pure de Sitter space is known to be a solution of Vasiliev's higher spin gravity with only the metric turned on and all other fields turned off. It corresponds to the CFT living on some three-geometry with all sources turned off.  Which three-geometry depends on how the boundary is approached.  Some common choices are the standard metrics on $S^3$, $\mathcal{H}^3$, $\mathbb{R}^3$ and $S^2\times \mathbb{R}^1$. They correspond to the global, hyperbolic (relevant to the nucleation of a Coleman-de Luccia bubble), planar and static patch slicings of pure de Sitter space. 

In addition, one may consider quotients of the above geometries which are still solutions in Vasiliev gravity. Such quotients can make the de Sitter slicing compact.  The simplest example is the $r_0=0$ solution we studied in section \ref{einstsec}, but there are many other options.  For instance, we can quotient $\mathbb{R}^3$ to  a $T^3$. This geometry shrinks to zero size in the infinite past, touching a point at $\mathcal{I}^-$. Another example corresponds to quotienting $\mathcal{H}^3$ to a compact manifold. In this case the universe shrinks to a point at a finite proper time in the past. This is clear from the metric of de Sitter space with $\mathcal{H}^3$ slices:
\begin{equation}
{ds^2} = {\ell_{dS}^2} \left( -d\tau^2 + \sinh^2 \tau \; d \mathcal{H}_3^2 \right)~,
\end{equation}
where $d \mathcal{H}_3^2 = \left( d\rho^2 + \cosh^2\rho \; d\Omega^2_2 \right)$ is the canonical unit metric on $\mathcal{H}^3$. The `big bang' would occur at $\tau=0$. 

From the perspective of the dual CFT, we are considering the $Sp(N)$ theory on a compact quotient of $\mathcal{H}^3$. Given that we must impose a singlet constraint, it is rather complicated to actually compute the partition function as a function of the geometric data of the quotient. On the other hand, it is well known mathematically that such compact quotients have a rich topological structure which the integral over the space of flat connections will be sensitive to. It would be interesting to understand more deeply how the presence of a `big bang' is connected to this topological structure.

We should mention a criticism to the above. Vasiliev theories are non-local for distances smaller than the de Sitter length. This is due to an infinite sequence of higher derivative terms in the equations of motion and the fact that the de Sitter length is the only scale in the theory. The big bang singularity we are discussing is at scales far smaller than the de Sitter scale, so without the aid of a good gauge invariant observable we cannot conclude whether there indeed exists a pathology. On a more perturbative level however, these solutions have a large metric turned on and all other fields turned off, thus a probe like observer would indeed perceive a shrinking universe. Some insight may be gained from a linear analysis about one of these `big bang' cosmologies.

%

\section{Outlook}\label{outlook}

The nature of our work has been explorative. There are many directions in which we must proceed to flesh out our understanding of de Sitter holography, even in the context of the simple example we are studying. It would be interesting to understand how much of our picture is qualitatively similar to the case of more ordinary theories of Einstein gravity. An example of this is the divergence at small $S^1$ relative to $S^2$ discussed in section \ref{s2s1sect}. 

An interesting question that naturally arises is that of topology. How are we to compare the wave function evaluated on different topologies? In situations with more complicated topologies the Chern-Simons piece of the action will play a significant role.\footnote{An obvious guess for how to compare different topologies is to normalize the Chern-Simons integration measure on some manifold $\mathcal{M}$ as in \cite{Witten:1988hf}. In this normalization the pure Chern-Simons partition function on $\mathcal{M} = S^2 \times S^1$ is $Z_{CS}\left[S^2 \times S^1\right] = 1$ for any gauge group and $k$, whereas it vanishes at large $k$ on an $S^3$.  More generally this normalization would suggest that more sophisticated topologies are infinitely preferred as $k\to\infty$.  We are unsure however if this is really the correct way to compare topologies.} A simple example that seems calculable is the partition function on a Lens space (i.e. a quotient of $S^3$ by the cyclic group of order $p$,  $\mathbb{Z}_p$, which has finite fundamental group). Such Lens spaces provide a discretum of topologies that we can study the wave function on. More generally, inspired by \cite{shenkermaltz}, we might also envision studying the wave function on more general higher genus surfaces. 

We must also confront the issue of turning on sources for the higher spin currents.  One approach would be to sort out the higher spin spherical harmonics needed to complete the argument of section \ref{normal} and decisively confirm the non-normalizeability of the wave function.  Alternatively, given that the higher spin currents are all {\it quadratic} in the fields $\chi^a$, one expects schematically that the action for general sources is given by the bilocal expression \cite{Douglas:2010rc,Das:2012dt}:
\begin{multline}
S_{bare}[B,g_{ij},\chi] = \frac{1}{2} \int d^3 x \sqrt{g} \Omega_{ab} \left[ \partial_i \chi^a \partial_j \chi^b g^{ij} + \frac{1}{8} R[g] \chi^a\chi^b \right] + \\  \int d^3x \sqrt{g(x)} \int d^3 y \sqrt{g(y)} \Omega_{ab} \chi^a (x) B(x,y) \chi^b (y)~.
\end{multline}
The challenge of course is to understand what we mean be the space of all $B(x,y)$'s and how to relate the partition function $Z_{CFT} = \int \mathcal{D} \chi \exp(-S_{bare}[B,g_{ij},\chi])$ back to a wave function whose coordinates are the sources of each current, but this formalism may simplify the treatment of the higher spin sources.

Finally if the scalar divergence indeed indicates an instability of Vasiliev's theory in de Sitter space, it would be very interesting to understand the bulk mechanism by which this decay proceeds.  In particular we would like to know what is the fate of Vasiliev's universe; where does it decay to?  It has been recently argued \cite{Chang:2012kt} that the AdS version of Vasiliev's higher-spin theories can be realized as various limits of string theory; perhaps these decays connect higher-spin theories dynamically to the rest of the string landscape \cite{Bousso:2000xa,Kachru:2003aw,Susskind:2003kw}.

\section*{Acknowledgements}
It was a great pleasure discussing this work with Tarek Anous, Nicolas Boulanger, Alejandra Castro, George Coss, Xi Dong, Jim Hartle, Tom Hartman, Sean Hartnoll, Simeon Hellerman, Thomas Hertog, Daniel Jafferis, Alex Maloney, Rob Myers, Edgar Shaghoulian, Steve Shenker, Eva Silverstein, Douglas Stanford, Andy Strominger, Lenny Susskind, Per Sundell, Misha Vasiliev, and Xi Yin. The authors are also grateful to the ``Cosmology and Complexity" conference at Hydra for their hospitality while this work was in progress. This work has been partially funded by DOE grant DE-FG02-91ER40654 and by a grant of the John Templeton Foundation. DH is funded by the Stanford Institute for Theoretical Physics and NSF Grant 0756174.  The opinions expressed in this publication are those of the authors and do not necessarily reflect the views of the John Templeton Foundation.

\appendix
\section{Double Trace Flows in dS/CFT}\label{altquantapp}

In AdS/CFT it has long been understood \cite{Klebanov:1999tb} that there are two ``natural'' boundary conditions for a bulk scalar field, which are usually referred to as the ``standard'' and ``alternate'' quantizations.  The partition functions for the two boundary conditions are related by a double-trace RG flow \cite{Gubser:2002vv}.  In dS/CFT the analogue of the standard quantization of a bulk scalar is the Hartle-Hawking wave function in field space, while the analogue of the alternate quantization is the \textit{same} wave function projected onto a different basis of the Hilbert space of field configurations.  In this appendix we develop this connection on the field theory side, showing in more detail how it arises from a double-trace RG flow.  We will give arguments valid for a general large-$N$ CFT in $d$ dimensions, which we will specialize to the Vasiliev case theory at the end.

In AdS/CFT the observation that boundary conditions in the bulk are a statement about states in the radial bulk Hilbert space goes back at least to \cite{Witten:2001ua}; our discussion here builds primarily on \cite{Gubser:2002vv} and \cite{Hartman:2006dy}.  These issues were also recently studied in the $O(N)$ model in \cite{Giombi:2011ya}, whose results we use to resolve an important technical subtlety. 

\subsection{Field Theory Double-Trace Flow}
Consider a CFT on $\mathbb{R}^d$ that is deformed by a scalar operator $\mathcal{O}$ having dimension $\Delta<d/2$:
\be\label{altpart}
Z_{alt}[\sigma]=\int \mathcal{D}\mathcal{M}e^{-S_{CFT}+\int d^d x \sigma(x)\mathcal{O}(x)}~.
\ee
For technical reasons we will also assume that $\Delta>\max\left(\frac{d}{3},\frac{d-2}{2}\right)$.\footnote{This inequality is actually saturated for the $Sp(N)$ model; we comment on this below.}  Say that the lowest dimension non-identity operator appearing in the $\mathcal{O}\mathcal{O}$ operator product expansion is a scalar, which we denote $\mathcal{O}^2$, and that there is some parameter $N$ such that its scaling dimension is $\Delta_{2}=2\Delta+O(1/N)$.  We will normalize $\mathcal{O}$ so that its two-point function is $O(N)$.\footnote{We have defined $N$ in a way that suggests a vector theory, but this section is also valid for matrix theories provided one replaces $N\to N^2$.}  It is widely believed \cite{Witten:2001ua,Gubser:2002vv} that at finite but large enough $N$ the deformed theory
\be
Z_{flow}[0]= \int \mathcal{D}\mathcal{M}e^{-S_{CFT}-\frac{f}{2N}\int d^d x \mathcal{O}^2}
\ee  
flows to an IR fixed point at which the dimension of $\mathcal{O}$ becomes $\tilde{\Delta}=d-\Delta + O(1/N)$.  The scale below which the theory is near the IR fixed point is determined by the dimensionful parameter $f$, so by choosing $f$ to be of order the cutoff length to the appropriate power we can directly study the IR fixed point.  To be definite we will set
\be
f\equiv f_0\epsilon^{\Delta_2-d}~,
\ee
with $f_0$ some dimensionless constant that is $O(N^0)$.  We can then study a scalar deformation of this fixed point by computing\footnote{In this equation and the previous one we should include a ``bare source'' proportional to $\mathcal{O}$ in the action, which we tune to ensure that the one point function of $\mathcal{O}$ is zero at the IR fixed point.  The coefficient of this term is zero to leading order in $N$, so rather than carry it around explicitly we instead declare that in what follows one should understand our $\tilde{\sigma}$ to differ from the ``true'' $\tilde{\sigma}$ by an additive constant that vanishes at large $N$.  In equation \eqref{appdSBC} below this translates into a small constant additive shift of $\sigma$; this has no meaningful effect on the physics of the main text so we ignore it from here out.}
\be\label{stanCFT}
Z_{stan}[\tilde{\sigma}]=e^{-S_{ct}[\tilde{\sigma}]}\int \mathcal{D}\mathcal{M}\exp\left[-S_{CFT}+\rho f \epsilon^{\gamma}\tilde{\sigma}\mathcal{O}-\frac{f}{2N}\int d^d x \mathcal{O}^2\right]~.
\ee
Here $\gamma\equiv d+\Delta-\tilde{\Delta}-\Delta_2=O(1/N)$, which ensures that $\tilde{\sigma}$ has dimension $d-\tilde{\Delta}$ as expected from the dS/CFT dictionary.  $\rho$ is a dimensionless constant we introduce so that we may take $\tilde{\sigma}$ to be related to a canonically normalized bulk field in the standard way:
\be
\phi=\sqrt{N}\epsilon^{d-\tilde{\Delta}}\tilde{\sigma}~.
\ee 
The value of $\rho$ will of course depend on the CFT normalization of $\mathcal{O}$.  The counterterm action $S_{ct}$ is included to ensure that we get the full bulk wave function including local terms.  For the range of $\Delta$ we are studying there is only one counterterm that does not vanish as $\epsilon\to 0$, so we simply have
\be
S_{ct}=\frac{1}{2}\alpha N\epsilon^{d-2\tilde{\Delta}}\int d^d x \tilde{\sigma}^2~,
\ee
with $\alpha$ a tuneable dimensionless parameter we determine below by comparison with the bulk.

Following \cite{Gubser:2002vv} we observe that that $Z_{stan}$ and $Z_{alt}$ are related by Hubbard-Stratonovich transformations.  To simplify formulas we will from here on neglect $O(1/N)$ corrections in powers of the cutoff; these can be restored at any point by dimensional analysis.\footnote{We will also assume that we can freely multiply $\mathcal{O}$'s together using $\mathcal{O}\times\mathcal{O}=\epsilon^{\Delta_2-2\Delta}\mathcal{O}^2$.  We should in principle also include the UV-divergent identity contribution to $\mathcal{O}\mathcal{O}$, but this gives only scheme-dependent factors which are independent of $\sigma$ and $\tilde{\sigma}$ and thus can be absorbed into the normalization of $Z_{stan}$ and $Z_{alt}$ in a trivial way.}  We may then write:
\begin{align}\label{st2alt}
Z_{alt}[\sigma]&=\sqrt{\text{det}\left(\frac{\rho f N}{2\pi}\right)}e^{-\frac{N}{2f}\int d^d x\sigma^2}\int \mathcal{D}\tilde{\sigma}e^{S_{ct}[\tilde{\sigma}]+N\int d^d x \left(-\frac{\rho^2 f \tilde{\sigma}^2}{2}+\rho \sigma\tilde{\sigma}\right)}Z_{stan}[\tilde{\sigma}]\\\label{alt2st}
Z_{stan}[\tilde{\sigma}]&=\sqrt{\text{det}\left(-\frac{N}{2\pi f}\right)}e^{-S_{ct}[\tilde{\sigma}]+\frac{N f\rho^2}{2} \int d^d x \tilde{\sigma}^2}\int \mathcal{D}\sigma e^{N\int d^d x\left(\frac{\sigma^2}{2f}-\rho\sigma\tilde{\sigma}\right)}Z_{alt}[\sigma]~.
\end{align}
The integration contours depend on the phase of $f$ and are chosen to ensure convergence.  It was pointed out in \cite{Gubser:2002vv} that in the limit $\epsilon \to 0$ (which here just means $f\to\infty$) the transformation \eqref{alt2st} becomes a functional Legendre transformation up to a local counterterm; the inverse transformation \eqref{st2alt} becomes singular in a nontrivial way.  In the following subsection we will interpret these transformations as a quantum mechanical change of basis.

\subsection{Bulk Boundary Conditions}

In this section we recall some basic properties of the wave function of a massive scalar in dS space.  For simplicity we will use the formalism of a scalar field in a fixed $dS$ background, but we could also make the argument in Wheeler-deWitt language.  We will use coordinates where
\be
ds^2=\frac{1}{T^2}\left(-dT^2+d\vec{x}^2\right)~.
\ee
For a scalar action of the form
\be
S_{bulk}=-\int d^{d+1}\sqrt{g}\left[\frac{1}{2}\partial_\mu \phi \partial^\mu \phi +V(\phi)\right]~,
\ee
with $T<0$, we can discretize the bulk path integral expression of the Bunch-Davies wave function to find
\begin{align}\nonumber
\Psi[\phi,T+\delta]=&\sqrt{\frac{-i(-T)^{1-d}}{2\pi\delta}}\int \mathcal{D}\hat{\phi}\\\label{dsPsi}
&\exp\left[i\int d^d x\left\{\frac{(-T)^{1-d}}{2\delta}(\phi-\hat{\phi})^2-\delta (-T)^{-(d+1)}\left(V(\hat{\phi})+\frac{T^2}{2}\partial_i\tilde{\phi}\partial_i\hat{\phi}\right)\right\}\right]\Psi[\hat{\phi},T]~,
\end{align}
to linear order in $\delta$.  Expanding both sides in $\delta$ one can derive the functional Schrodinger equation for $\Psi$.  We can easily see that canonical momentum conjugate to $\phi$ is
\be
\Pi\equiv -i\frac{\delta}{\delta\phi}=(-T)^{1-d}\partial_T \phi~.
\ee
Our introduction of $S_{ct}$ and $\rho$ in the previous section allow a simple dS/CFT bulk-boundary dictionary:
\be
Z_{stan}[\tilde{\sigma}]=\Psi[\sqrt{N}\tilde{\sigma}\epsilon^{d-\tilde{\Delta}},-\epsilon]~.
\ee
Using this dictionary we can apply the transformation \eqref{st2alt} to the \eqref{dsPsi} to find a bulk interpretation for $Z_{alt}$.  The integral over $\tilde{\sigma}$ is Gaussian and can be done exactly - it imposes boundary conditions
\footnote{As mentioned in a previous footnote, because the $Sp(N)$ model saturates the allowed region for $\Delta$ it is in principle possible to have an additional term  in $S_{ct}$ proportional to $\tilde{\sigma}^3$.  This would lead to an additional term proportional to $N^{-1} \phi^2$ on the left hand side of \eqref{appdSBC}.  This term would be of importance to us since we are interested in field configurations where $\phi\sim \sqrt{N}$.  The non-existence of such a term can be inferred from the work of \cite{Giombi:2011ya}, who showed order by order in $1/N$ that the critical and free $O(N)$ models are related in the same way as the bulk correlation functions built from propagators with Dirichlet boundary conditions or \eqref{appdSBC} respectively.*  This agreement would be spoiled by a quadratic term in the boundary conditions.  The argument of \cite{Giombi:2011ya} used special properties of the $O(N)$ model, which carry over straightforwardly to the $Sp(N)$ model.  It is worth mentioning that the interpretation of this term, according to the general proposal for multitrace boundary conditions in \cite{Witten:2001ua}, is that were it included it would deform the free $Sp(N)$ model by the approximately marginal operator $(\chi^a\Omega_{ab}\chi^b)^3$.

*Actually \cite{Giombi:2011ya} don't explicitly mention \eqref{appdSBC}, the detailed equivalence of the ``alternate'' propagators they use to \eqref{appdSBC} for the AdS case was shown earlier in \cite{Hartman:2006dy}.}
\be\label{appdSBC}
-T \partial_T \phi+i(\rho^2f_0-\alpha)\phi=i\rho \sqrt{N}\sigma (-T)^{d-\Delta}~.
\ee
Thus $Z_{alt}$ is the same bulk path integral but with different boundary conditions.  For real field configurations the left hand side of \eqref{appdSBC} should be real.  As we noted before the value of $\rho$ will depend on the choice of normalization of $\mathcal{O}$ - we here set the phase of $\langle \mathcal{O}\mathcal{O}\rangle_{alt}$ such that this reality requires $\sigma$ to be real.  We then see that $\rho$ and $(\rho^2 f_0-\alpha)$ must be purely imaginary.  

With this phase choice we can then interpret $Z_{alt}$ as the Hartle-Hawking wave function projected onto eigenstates of the hermitian operator
\be
\Sigma=\frac{-i(-T)^{\Delta-d}}{\rho \sqrt{N}}\left(-(-T)^{d} \Pi+i(\rho^2f_0-\alpha)\phi\right)~.
\ee

Note that if we had not imposed a lower bound on $\Delta$, there could have been additional terms in $S_{ct}$ and the boundary conditions could be more complicated.

\subsection{Explicit Parameters at large N}
At large $N$ we can determine $\rho$ and $\alpha$ by comparison of  $\langle\mathcal{O}\mathcal{O}\rangle_{stan}$ to a free bulk scalar computation of $\Psi$.  The needed bulk result is \cite{Harlow:2011ke}
\be\label{bulkdSmassive}
\Psi[\phi,T]=\exp \left[i\frac{\tilde{\Delta}-d}{2(-T)^d}\int \frac{d^dk}{(2\pi)^d}\phi_{-k}\phi_k\left(1-c e^{\frac{i\pi}{2}(2\tilde{\Delta}-d)}(-kT)^{2\tilde{\Delta}-d}+\ldots \right)\right]~,
\ee
where $c=\frac{2^{d-2\tilde{\Delta}+1}}{d-\tilde{\Delta}}\frac{\Gamma(d/2-\tilde{\Delta}+1)}{\Gamma(\tilde{\Delta}-d/2)}$ and $\ldots$ means terms that are higher order in $-T$.  $\tilde{\Delta}$ is related to the bulk mass in the usual dS/CFT way:
\be
\tilde{\Delta}=\frac{d}{2}+\frac{1}{2}\sqrt{d^2-4m^2}~.
\ee

We can compare this with the standard large $N$ result \cite{Gubser:2002vv}
\be
\langle\mathcal{O}_k\mathcal{O}_{k'}\rangle_{stan}=\delta^d(k+k')N\epsilon^{d-2\Delta}\left(\frac{1}{f_0}-\frac{c'}{f_0^2}(k\epsilon)^{d-2\Delta}+\ldots\right)~;
\ee
if we define 
\be
\langle\mathcal{O}(x)\mathcal{O}(y)\rangle_{alt}=\frac{N\mathcal{C}}{|x-y|^{2\Delta}}
\ee
then we have \cite{Gubser:2002vv}
\be
c'=\mathcal{C}^{-1}2^{2\Delta-d}\pi^{-d/2}\frac{\Gamma(\Delta)}{\Gamma(d/2-\Delta)}~.
\ee
By matching the second derivative of $Z_{stan}$ to \eqref{bulkdSmassive} we then find
\be\label{rho2f}
\alpha-\rho^2 f_0=i \Delta+O(1/N)~,
\ee
and
\be
\rho^2=-2i\pi^{\frac{d}{2}}\frac{\Gamma(\Delta-\frac{d}{2}+1)}{\Gamma(\Delta)}e^{\frac{i\pi}{2}(d-2\Delta)}\mathcal{C}+O(1/N)~.
\ee

Finally for use in the main text we note that in the $Sp(N)$ model we have $d=3$, $\Delta=1$, and if we define 
\be
\mathcal{O}=J^{(0)}=-\frac{1}{2}\Omega_{ab}\chi^a\chi^b~,
\ee
then $\mathcal{C}=-\frac{1}{32\pi^2}$.  This gives $\rho^2= -1/16+O(1/N)$ and $\alpha-f_0\rho^2=i+O(1/N)$, so the boundary conditions become\footnote{Here we made the arbitrary sign choice $\rho=i/4$.  This amounts to choosing the sign of the bulk scalar field.}
\be
T\partial_T \phi+\phi=\frac{1}{4}\sqrt{N}T^2\sigma~.
\ee  
In fact for the $Sp(N)$ model the $O(1/N)$ corrections to $\rho$ and $\alpha$ actually are zero; this follows from the argument of \cite{Giombi:2011ya}, and uses special properties of the $Sp(N)$ model.  This cancellation is not necessary for the general interpretation of the alternate quantization, but it makes things more convenient in the main text so we will use it.    

\end{document}